%% file: main.tex
\documentclass[sn-basic]{sn-jnl}

\input{preprocessor.tex}

\jyear{2022}%

\begin{document}

\title[Foveated Computational Meta Optics]{Foveated Thermal Computational Imaging\\ in the Wild Using All-Silicon Meta-Optics}


\author[1]{\fnm{Vishwanath} \sur{Saragadam}}\email{vishwanath.saragadam@rice.edu}

\author[2]{\fnm{Zheyi} \sur{Han}}\email{zh25@uw.edu}

\author[1]{\fnm{Vivek} \sur{Boominathan}}\email{vivekb@rice.edu}

\author[2]{\fnm{Luocheng} \sur{Huang}}\email{luocheng@uw.edu}

\author[1]{\fnm{Shiyu} \sur{Tan}}\email{st64@rice.edu}

\author[2,3]{\fnm{Johannes E.} \sur{Fr\"och}}\email{jfroech@uw.edu}

\author[2,4]{\fnm{Karl~F.} \sur{B\"ohringer}}\email{karlb@uw.edu}

\author[1]{\fnm{Richard} \sur{G.~Baraniuk}}\email{richb@rice.edu}

\author[2,3]{\fnm{Arka} \sur{Majumdar}}\email{arka@ece.uw.edu}

\author*[1]{\fnm{Ashok} \sur{Veeraraghavan}}\email{vashok@rice.edu}

\affil*[1]{\orgdiv{Dept. of ECE}, \orgname{Rice University}, \orgaddress{\city{Houston}, \postcode{77005}, \state{TX}, \country{USA}}}

\affil[2]{\orgdiv{Dept. of ECE}, \orgname{University of Washington}, \orgaddress{\city{Seattle}, \postcode{98195}, \state{WA}, \country{USA}}}

\affil[3]{\orgdiv{Dept. of Physics}, \orgname{University of Washington}, \orgaddress{\city{Seattle}, \postcode{98195}, \state{WA}, \country{USA}}}

\affil[4]{\orgdiv{Institute for Nano-Engineered Systems}, \orgname{University of Washington}, \orgaddress{\city{Seattle}, \postcode{98195}, \state{WA}, \country{USA}}}



\abstract{Foveated imaging provides a better tradeoff between situational awareness (field of view) and resolution, and is critical in long wavelength infrared regimes because of the size, weight, power, and cost of thermal sensors.  We demonstrate computational foveated imaging by exploiting the ability of a meta-optical frontend to discriminate between different polarization states and a computational backend to reconstruct the captured image/video. The frontend is a three-element optic: the first element which we call the ``foveal" element is a metalens that focuses s-polarized light at a distance of $f_1$ without affecting the p-polarized light; the second element which we call the ``perifoveal" element is another metalens that focuses p-polarized light at a distance of $f_2$ without affecting the s-polarized light. The third element is a freely rotating polarizer that dynamically changes the mixing ratios between the two polarization states. Both the foveal element (focal length = 150mm; diameter = 75mm), and the perifoveal element (focal length = 25mm; diameter = 25mm) were fabricated as polarization-sensitive, all-silicon, meta surfaces resulting in a large-aperture, 1:6 foveal expansion, thermal imaging capability. A computational backend then utilizes a deep image prior to separate the resultant multiplexed image or video into a foveated image consisting of a high-resolution center and a lower-resolution large field of view context. We build a first-of-its-kind prototype system and demonstrate 12 frames per second real-time, thermal, foveated image and video capture in the wild.}


\hyphenation{meta-lens}
\hyphenation{meta-lenses}



\maketitle

\section{Introduction}\label{sec1}
\input{intro.tex}
\section{Results}\label{sec2}
\input{results.tex}

\section{Discussion}\label{sec3}
\input{disc.tex}

\section{Materials and Methods}\label{sec4}
\input{methods.tex}

\backmatter
%
%

\bmhead{Acknowledgments}
The work is supported by NSF grants NSF-2127235, CCF-1911094, IIS-1838177, IIS-1730574, IIS-1652633, and IIS-2107313; ONR grants N00014-18-12571, N00014-20-1-2534, and MURI N00014-20-1-2787; AFOSR grant FA9550-22-1-0060; and a Vannevar Bush Faculty Fellowship, ONR grant N00014-18-1-2047. Part of this work was conducted at the Washington Nanofabrication Facility / Molecular Analysis Facility, a National Nanotechnology Coordinated Infrastructure (NNCI) site at the University of Washington with partial support from the National Science Foundation via awards NNCI-1542101 and NNCI-2025489.

%





\begin{appendices}
\section{Computational Reconstruction}\label{sec:appendix1}
\input{sup_algo.tex}

\section{Experimental Details}\label{sec:appendix2}
\input{sup_experiments.tex}

%




\end{appendices}


\bibliography{refs_main, refs_sup}


\end{document}

%% file: preprocessor.tex
\usepackage{graphicx}
\usepackage{amsmath}
\usepackage[T1]{fontenc}
\usepackage{lmodern}

\usepackage{amssymb}
\usepackage{subcaption}
\usepackage{enumitem}
\usepackage{comment}
\usepackage{url}
\usepackage{xcolor}
\usepackage{capt-of}
\usepackage{etoolbox}
\usepackage[utf8]{inputenc} 
\usepackage{hyperref}       
\usepackage{url}            
\usepackage{booktabs}       
\usepackage{amsfonts}       
\usepackage{nicefrac}       
\usepackage{microtype}      
\usepackage{xcolor}         
\usepackage{balance}

\setlist{leftmargin=*}

\usepackage{algorithm}
\usepackage{algpseudocode}

\newcommand{\bfx}{{\bf x}}
\newcommand{\bfy}{{\bf y}}

\newcommand{\bfs}{{\bf s}}

\newcommand{\bfn}{{\bf n}}

\newcommand{\bpara}[1]{\vspace{0.5em}\noindent{\bf #1}}

%% file: intro.tex
\begin{figure}[!tt]
    \centering
    \includegraphics[width=\textwidth]{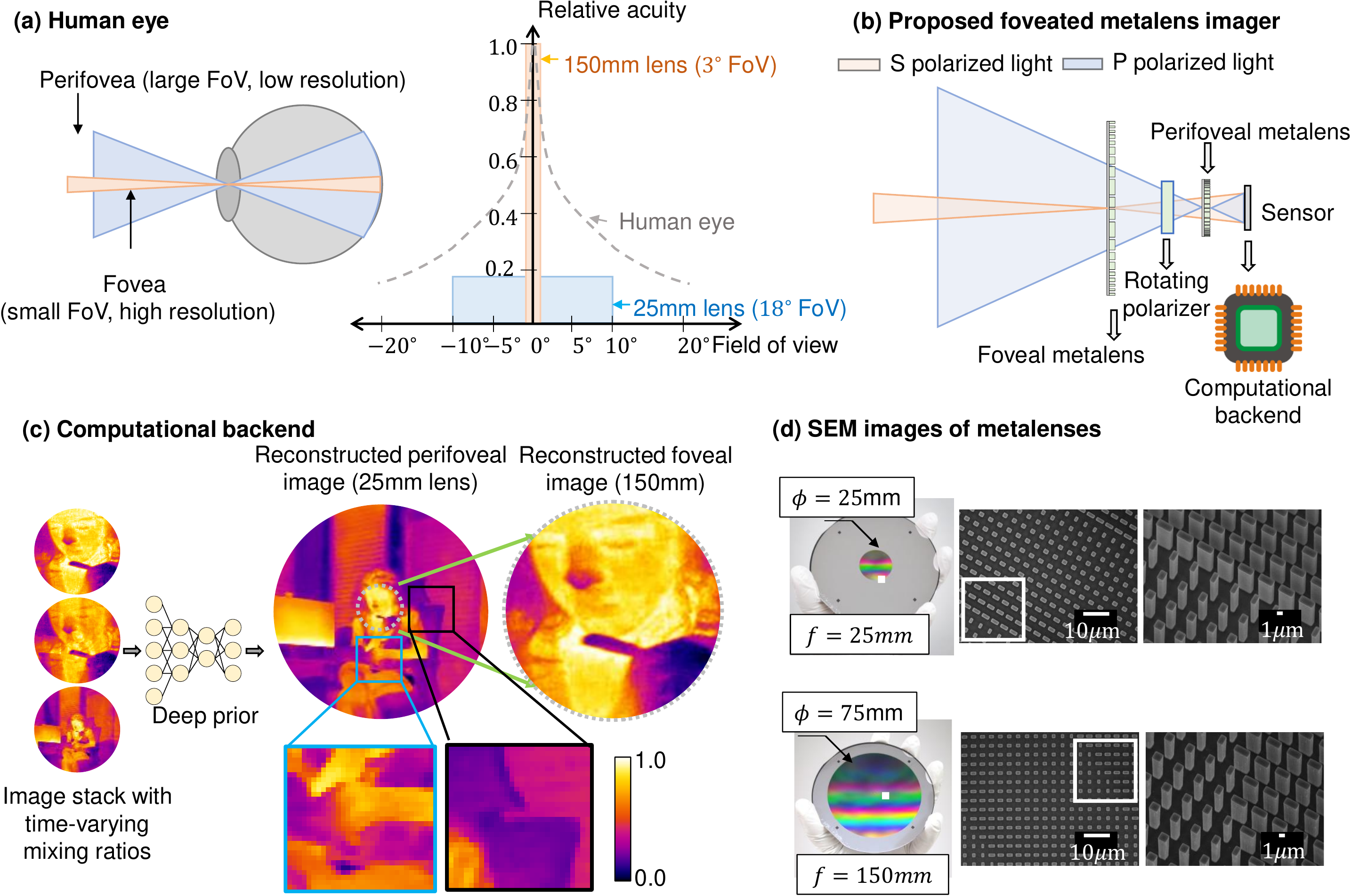}
    \caption{Overview of proposed foveated imaging system. (a) Our proposed foveated metalens is inspired by the human eye that has high acuity (high resolution) over a small field of view (fovea) and a low acuity over a large field of view (perifovea). (b) Our optic consists of a foveal element at a focal length of 150mm, a perifoveal element at 25mm, and a freely rotating polarizer. The foveal element only modulates the s-polarized light, while the perifoveal element only modulates the p-polarized light. The sensor measures a linear combination of the two images which is then unmixed with a computational backend. (c) Reconstruction with the computational backend for a simulated example. (d) Scanning electron microscopic images of the two lenses.}
    \label{fig:fig1}
\end{figure}

Nature has optimized the mammalian eye to have a large field of view (FoV) for situational awareness and high resolution in the central area for target detection with high precision.
In the human eye, the fovea (small central region covering a $2^\circ$ FoV) carries half as much information as the perifovea (large outer region covering more than $30^\circ$ FoV), thereby dedicating precious bandwidth to the central region (see Fig.~\ref{fig:fig1} (a)).
Such foveation has long been desired in artificial vision systems as well, since high resolution in the central area is desirable for target detection, while large FoV is essential for contextual awareness.
This is particularly true at long wave infrared (LWIR) wavelengths where detectors tend to be low-resolution making the simultaneous capture of high resolution target and large FoV context especially challenging.

Existing solutions for foveated imaging such as foveated sensors \cite{thiele20173d,carles2016multi} and foveating dynamic optics \cite{tremblay2013switchable} do not work for thermal imaging because of the unique challenges at those wavelengths.
Foveated sensor designs \cite{carles2016multi} are only practical when sensor pixel sizes are small. 
Unfortunately, thermal sensors have large pixel sizes due to their noise characteristics, making foveated sensor designs impractical.
Foveated dynamic optics such as \cite{tremblay2013switchable} have never been demonstrated in the thermal regime, primarily because of the challenges in dynamic control of optical properties in materials such as germanium and zinc selenide (materials used typically used to fabricate LWIR lenses).
In this paper, we leverage recent advances in two rapidly emerging fields -- meta-optics and computational imaging -- to demonstrate first-of-its-kind, real-time, thermal, foveated image and video capture in the wild.

Meta-optics with sub-wavelength feature sizes for arbitrary phase control~\cite{kamali2018review,zhan2017metasurface} have rapidly grown as alternatives to traditional optics~\cite{han2022millimeter,huang2021long,arbabi2018mems,vo2014sub,west2014all,lin2014dielectric,lu2010planar}, over a wide range of wavelengths and novel applications, including reflectors~\cite{arbabi2017planar,fan2017visible,fattal2010flat}, vortex beam generators~\cite{ran2018high,yue2016vector,ma2015planar,yang2014dielectric}, holographic masks~\cite{ren2020complex,hu2019trichromatic,zheng2015metasurface}, gratings~\cite{arbabi2020increasing,lalanne1998blazed}, optical convolutional neural networks~\cite{burgos2021design}, and polarization optics~\cite{dorrah2021metasurface,arbabi2015dielectric}.
Meta-optics have two distinct advantages over traditional optics: (a) multiplexing (combining multiple phase functions in a single surface) and (b) polarization control (meta-optics can be manufactured with different phase functions for orthogonal states of polarization), and these properties have been used previously for polarimetry~\cite{rubin2019matrix}, depth sensing~\cite{guo2019compact}, and tunable focus~\cite{tian2019dielectric,yao2021focusing}.

Here, we leverage the polarization sensitivity of meta-optics to design a three-element optic that achieves large-aperture foveated imaging at the LWIR wavelengths.
The first element which we call the ``foveal" element is a metalens that encodes a phase function for a convex lens of focal length $f_1$ for s-polarization, whereas the p-polarized light is transmitted without any change (or a lens with focal length $\infty$).
The second element which we call the ``perifoveal" elment is another metalens encodes a phase function for a convex lens of focal length $f_2$ for P-polarization, whereas the S-polarization that was modulated by the previous surface is left unaltered.
The third element is a freely rotating polarizer that dynamically changes the mixing ratios between the two polarization states.
This results in a linear expansion of the FoV by $f_1:f_2$ by the foveal lens, which we call the foveal expansion.
%
%
The schematic of the proposed optical system is shown in Fig.~\ref{fig:fig1}(b).
S and P polarized light independently focused by these two elements multiplexes on the image sensor, resulting in a video sequence that is not human-interpretable, but can be leveraged to reconstruct the two images with computational approaches. 

Over the last decade, computational imaging has emerged as a powerful tool, where signal processing~\cite{levin2007image,raskar2006coded,duarte2008single,sankaranarayanan2012cs,hitomi2011video,veeraraghavan2007dappled,zheng2013wide,wagadarikar2008single,saunders2019computational,shin2016photon} and machine learning \cite{baek2021single,tseng2021neural,li2020end,sitzmann2018end,chakrabarti2014rethinking} algorithms co-designed with such multiplexed imaging systems can undo the effects of multiplexing and reconstruct images and videos.
The freely rotating linear polarizer better-conditions the inverse problem by dynamically changes the fraction of S-polarized and P-polarized image that is multiplexed on the image sensor, meaning each frame in the captured video has a different weighted combination of the two images (one with foveal element and another with perifoveal element). 
The two images are then recovered by a computational backend that takes the sequence of images as input, and simultaneously estimates the time-varying mixing ratios, as well as the low-resolution, large FoV and high-resolution, narrow FoV images.
Most conventional deep learning based reconstruction algorithms require large amounts of in-domain training data to be successful, something that is challenging when using LWIR meta-optics due to poor signal to noise ratio (SNR) of LWIR sensors, as well as a paucity of training data. Our reconstruction algorithm consists of a deep generative prior~\cite{ulyanov2018deep} that has the distinct advantage of not requiring any training data, while producing high quality results. The pipeline for our computational backened as well as a simulated example is shown in Fig.~\ref{fig:fig1} (c).

We validated our approach by designing and fabricating the metalenses (optimized for $10\mu$m wavelength) in an all-silicon platform using direct laser writing!\cite{huang2021long}.
%
The foveal element was designed with a diameter of 75mm for a focal length of 150mm. The perifoveal element was designed with a diameter of 25mm for a focal length of 25mm. 
This resulted in a foveal expansion of $6:1$.
%
%
Images of the metalenses, as well as their scanning electron microscopic (SEM) images are shown in Fig.~\ref{fig:fig1} (d).
Our three-element optic demonstrates first-of-its-kind videos in the wild at real time ($>12$ frames per second) enabling high resolution thermal videos of buildings, moving cars, and dynamic humans.

%% file: results.tex
\bpara{System overview.} Figure~\ref{fig:fig1} shows an overview of the proposed foveated computational metaoptical system optimized for imaging at $10\mu m$ wavelength. 
The phase functions of the two metalenses are
\begin{align}
    &\left.\begin{array}{l}
    \phi_{1, P}(x, y) = \frac{2\pi}{\lambda_0}(\sqrt{x^2 + y^2 +f_1^2} - f_1)\\
    \phi_{1, S}(x, y) = \text{constant}\\
    \end{array}\right\}\text{Foveal metalens}\\
    &\left.\begin{array}{l}
    \phi_{2, P}(x, y) = \text{constant}\\
    \phi_{2, S}(x, y) = \frac{2\pi}{\lambda_0}(\sqrt{x^2 + y^2 + f_2^2} - f_2)
    \end{array}\right\}\text{Perifoveal metalens,}
\end{align}
where $\lambda_0 = 10\mu m$ is the design wavelength.
The foveal lens has a focal length of $f_1=150$mm and a diameter of 75mm, while the perifoveal lens has a focal length of $f_2=25$mm and a diameter of 25mm.
%
The lenses are placed at a distance of $f_1$ and $f_2$, respectively, from the sensor.
The linear polarizer rotating freely at 2Hz and is placed between the lens assembly and the sensor to capture measurements with varying intensity of the image from each metalenses. 
Additionally, we placed a 10$\mu$m spectral filter between the perifoveal metalens and the sensor for some of the experiments to reduce chromatic blur.
The sensor captures images 12 frames per second.


Let $I_1(x, y), I_2(x, y)$ be the images formed by the foveal and perifoveal metalenses respectively. Assuming the two images are static between $t=t_1$ and $t=t_2$, the resultant image on the sensor is
\begin{align}
	I_\text{meas.}(x, y) = \alpha_t I_1(x, y) + (1-\alpha_t)I_2(x, y),
\end{align}
where $\alpha_t = \cos^2(\theta_t)$ is the mixing ratio depending on the position at time $t$ of the linear polarizer at $\theta_t$.
We then estimate the images $I_1(x, y)$ and $I_2(x, y)$, and the mixing ratio $\alpha_t$ by solving the regularized linear inverse problem
\begin{align}
	\min_{\alpha_t, I_1(x, y), I_2(x, y)} \sum_{t=t_1}^{t_2}\left\| I_\text{meas.} -  \alpha_t I_1(x,y) - (1-\alpha_t) I_2(x,y)\right\|^2  \nonumber\\
	+\ \mathcal{R}( I_1(x,y), I_2(x,y)),
	\label{eq:optim}
\end{align}
where $\mathcal{R}$ is a regularization term that ensures spatial smoothness of individual images and encourages the two recovered images to be dissimilar.
Further details about the optimization procedure are provided in Section~\ref{sec4}.
%

%

\begin{figure}[!tt]
    \centering
    \includegraphics[width=\textwidth]{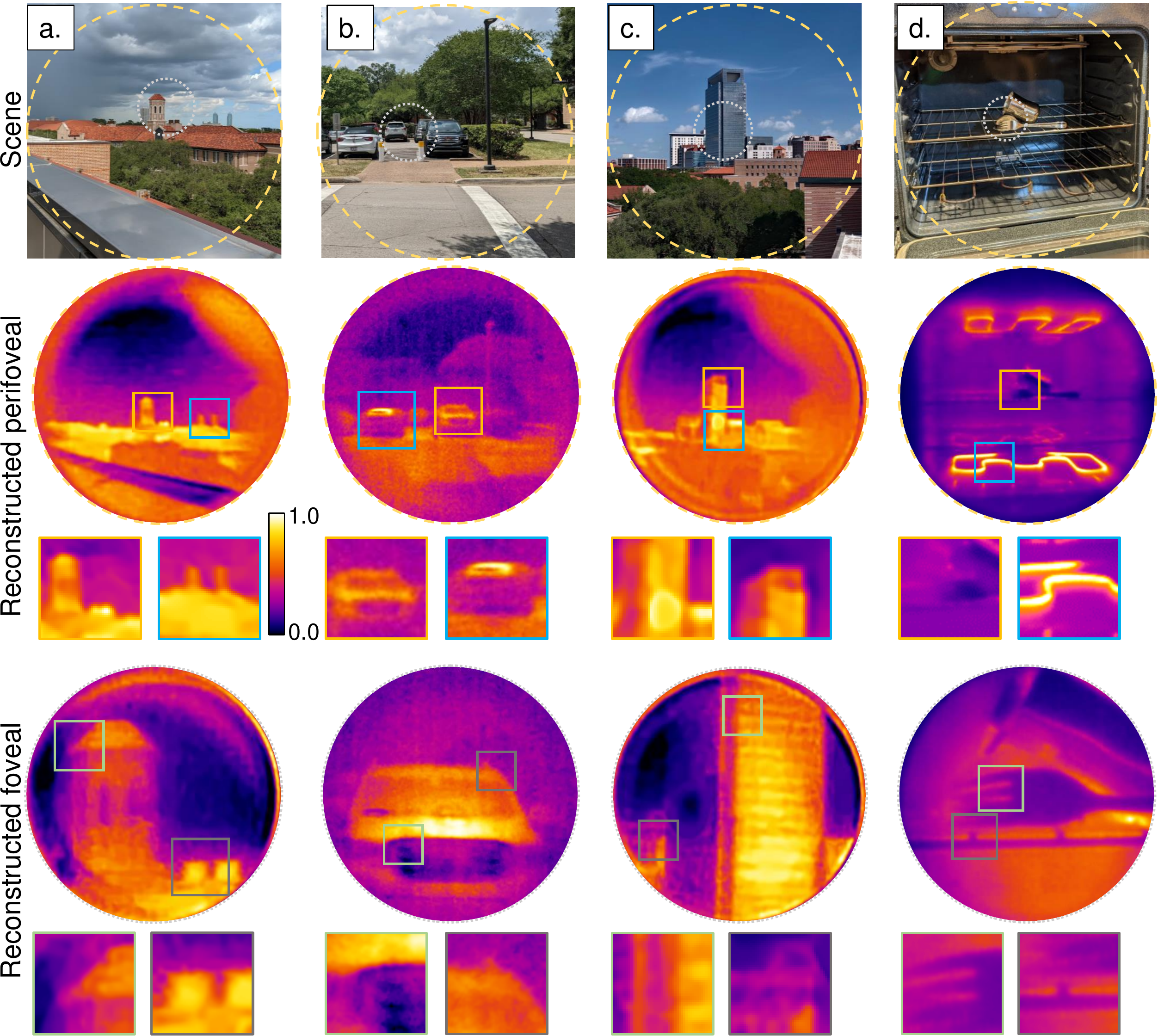}
    \caption{Static imaging in the wild. We captured images of several outdoor and indoor scenes including (a) rooftop, (b) cars in parking lot, (c) tall buildings, and (d) a hot oven. For scenes with objects at relatively lower temperature (a, c), we removed the narrowband spectral filter to increase SNR. Each of the scenes has a resolution of $290\times290$ pixels with each pixel measuring $24\mu m$. Our computational backend was able to recover very fine features such as windows in the rooftop building, the car grill, windows in the tall skyscraper, and the fork in the oven. All images use the same ``iron'' colorbar.}
    \label{fig:fig3}
\end{figure}

\bpara{Image reconstruction of still life in the wild.} 
Our optical setup produces high resolution foveated images in  diverse  indoor and outdoor settings.
The powerful computational backend is capable of reconstructing images from raw measurements in diverse settings. 
Figure~\ref{fig:fig3} shows reconstruction of several outdoor (a, b, c)  and indoors (d) scenes with a wide range of temperature variations.
In each case, 8 seconds of video data was captured with the polarizer rotating freely. 
We removed the spectral filter for Fig.~\ref{fig:fig3} (a, c) to increase measurement SNR.
We note that, while our metalens is designed only for 10-micron, we can still capture images under broadband light as demonstrated recently~\cite{huang2021long}.
The images are shown in the ``iron'' color map as is customary in thermal image visualization.
%
%
The advantages of dual focal length are evident -- the perifoveal image provides a context of the surroundings, while the foveal image provides high quality details of the central region. In particular, note the clear grating-like structure in Fig.~\ref{fig:fig3} (c) due to window frames, as well as the parallel prongs of the fork in Fig.~\ref{fig:fig3} (d).

\begin{figure}[!tt]
    \centering
    \includegraphics[width=\textwidth]{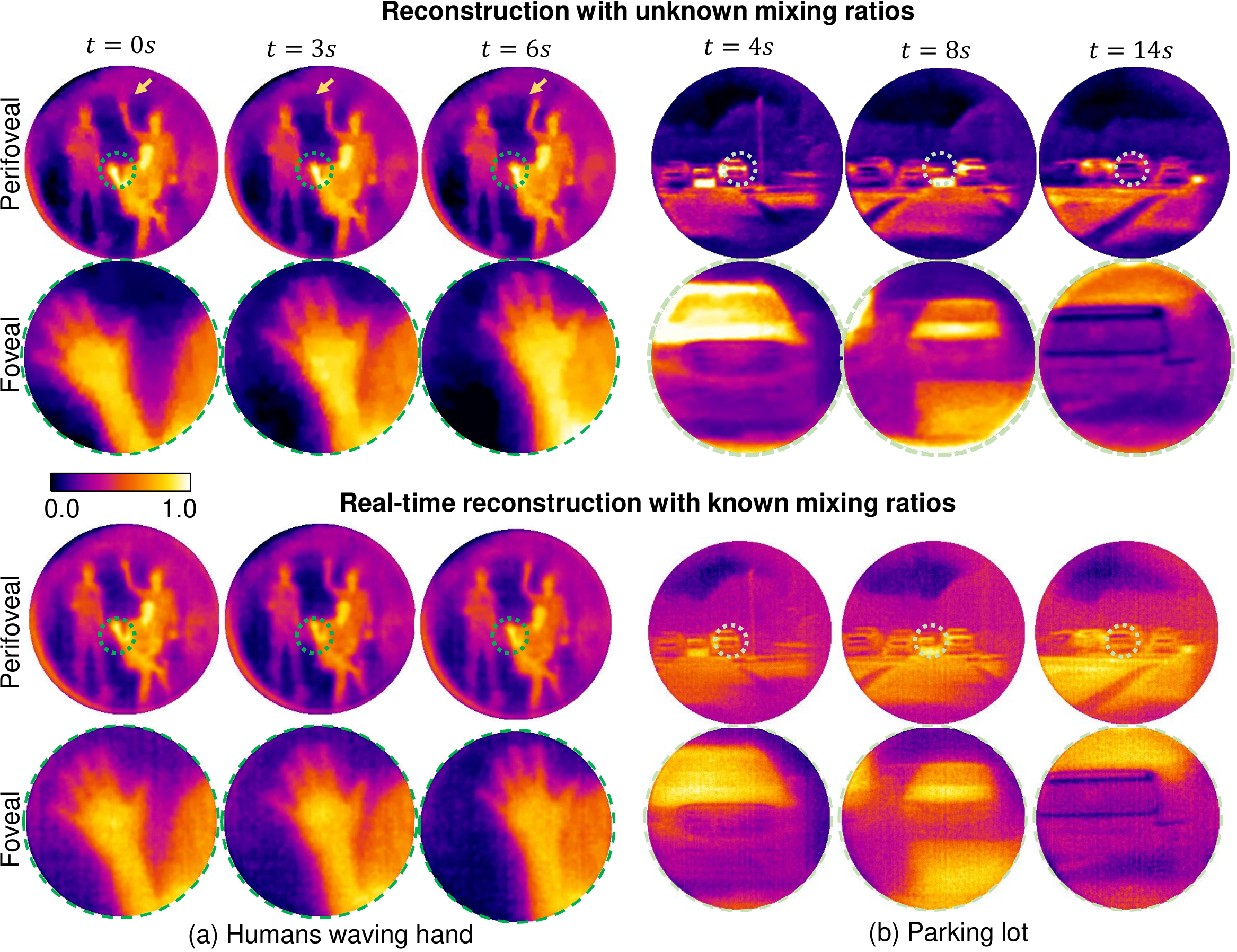}
    \caption{Capturing videos of dynamic scenes. Our optical setup enables simultaneous reconstruction of small and large FoV images of dynamic scenes with high accuracy. We leveraged a sliding window reconstruction where we assume that the scene does not change over eight frames (half the period of the polarizer) and then reconstruct a frame. Each of the scene has a resolution of $290\times290$ pixels with each pixel measuring $24\mu m$. Here we show reconstructed images from two videos with dynamic motion including (a) humans waving their hands, and (b) a  parking lot with several cars. If the mixing ratios are known apriori, then we can employ a simple least squares-based solution which enables reconstruction in real-time, shown in the third and fourth rows.}
    \label{fig:fig4}
\end{figure}
\bpara{Video reconstruction of dynamic scenes at 12 fps.} Our optical setup is capable of capturing dynamic scenes at high spatial and temporal resolutions, a result previously unreported in the metalens literature.
To demonstrate this, we imaged three people exhibiting different forms of motion, as shown in Fig.~\ref{fig:fig4} (a). The person on the left was rocking left to right, the person in the middle was sitting on a chair and waving his hand with small range of motion, and the person on the right was waving his hand with a large range.
%
%
We reconstructed a video sequence in a sliding window manner, where a continuous sequence of 8 measured images was used to recover a pair of perifoveal and foveal video frames.
Note the perifoveal image clearly showing waving of the right person's hand, but nearly no motion of the person sitting in the center.
The foveal however clearly shows the waving motion, along with all fingers separately.
Fig.~\ref{fig:fig4} (b) shows a parking lot with several cars. The perifoveal images capture the complete scene with all three cars, while the foveal images show details of each car in each snapshot.
If the mixing ratios $\alpha_t$ are known apriori, such as in the case when the polarizer rotates synchronously with the sensor exposure duration, then we can leverage a least squares-based reconstruction (details in section~\ref{sec:appendix2}) without any deep prior. This enables a real-time reconstruction, visualized in the second row in Fig.~\ref{fig:fig4}.
The computational time per frame was less than 20ms of CPU time per each frame.
This computation can be performed asynchronously while the camera is capturing images, thereby enabling a real-time reconstruction at 12 fps.
Thanks to the low computational complexity of OpenCV~\cite{opencv_library} functions, the computational backend can be implemented on a low-power computational platform such as a Raspberry-Pi, creating a compact imaging system.

\begin{figure}[!tt]
    \centering
    \includegraphics[width=\textwidth]{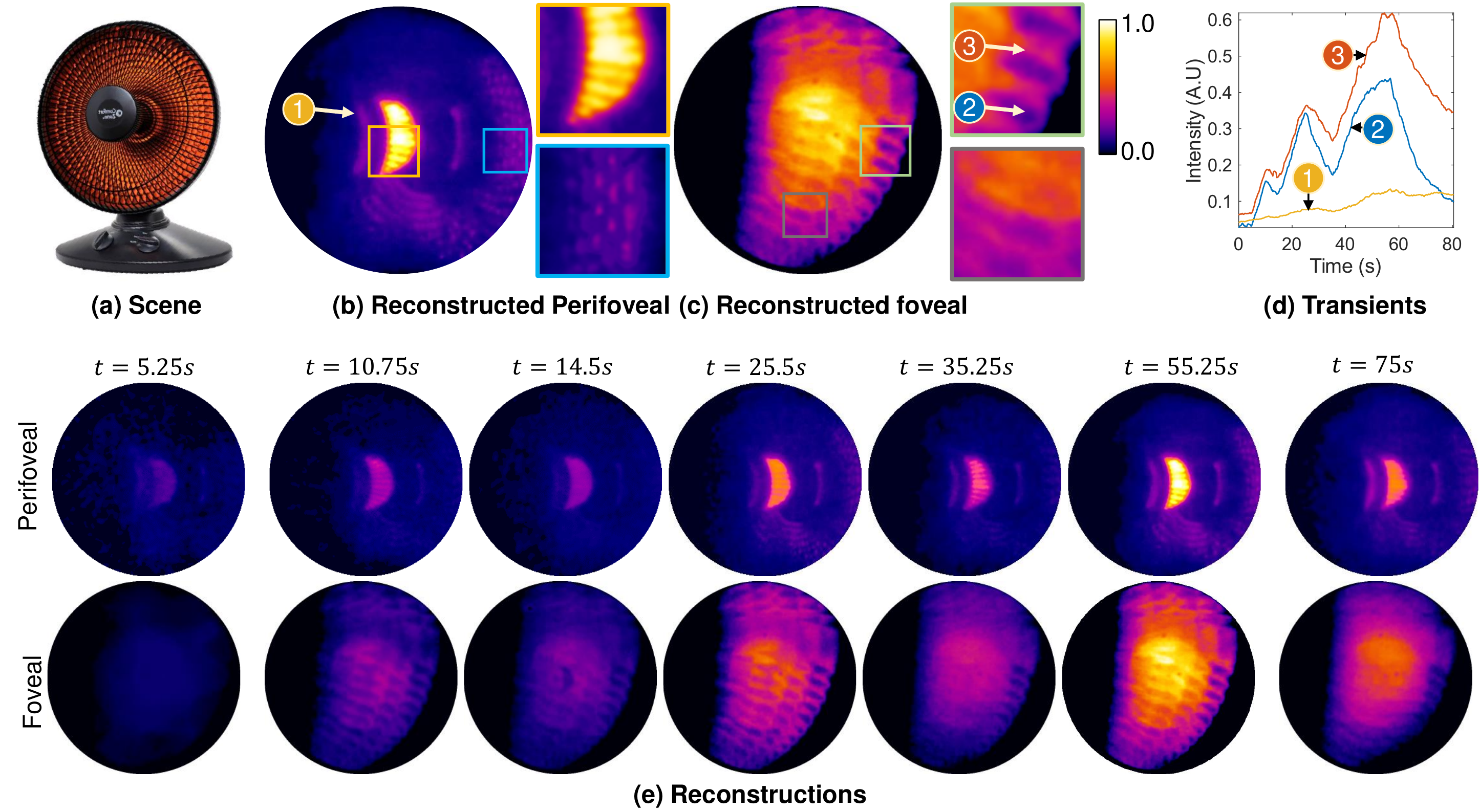}
    \caption{Imaging thermodynamics at high spatial and spectral resolutions. We switched a coil-based heater (a) on for 5s, then switched off for 5s, on for 10s, off for 10s, on for 20s and finally switched off in sequence, resulting in a heating and cooling periods, and then reconstructed (b) perifoveal and (c) foveal images. We can observe the complete heater, including the high frequency components of the parabolic reflector in the perifoveal images, as well as the coil in the foveal images. (d) transients over three marked areas show (1) indirect heating of the heating hub, (2) indirect heating close to the coil, and (3) direct heating of the coil. }
    \label{fig:fig5}
\end{figure}

As with dynamic motion, our setup can also image thermodynamic phenomena at high resolution. 
Figure~\ref{fig:fig5} (a) shows a portable heater going through three cycles of switching on and off (5s, 10s, and 20s), producing a distinct transient visualized in Fig.~\ref{fig:fig5} (d).
The effect of this heating cycle shows up in three forms. At the spatial point 1 that is indirectly heated, the rise and fall in relative intensity is small, and smooth. At point 2 that is also indirectly heated but close to the heating coil, the transients are stronger than at point 1. At point 3 on the coil, the transients are the strongest. 
Point 1 is easy to distinguish in the perifoveal image that provides an overall view of the heater, including the high frequency components of the parabolic reflector, but does not resolve the fine coil structure. In contrast, the foveal image clearly shows the coils and enables evaluation of transients at very fine-grained spatial resolutions.

\bpara{Single image reconstruction.}
While polarization control enables high quality image separation, it may not always be feasible to add a polarizer in the optical system, such as imaging with very high frame rates.
Our deep prior-based computational backend is sufficiently powerful to work in such scenarios, where we can only capture a single frame with a combination of images from both lenses.
There, we leverage the similarity between the downsampled foveal image, and the central crop of the perifoveal image to regularize the inverse problem.
Details about solving the inverse problem with a single measurement is provided in section~\ref{sec:appendix2}.

\begin{figure}[!tt]
	\centering
	\includegraphics[width=\textwidth]{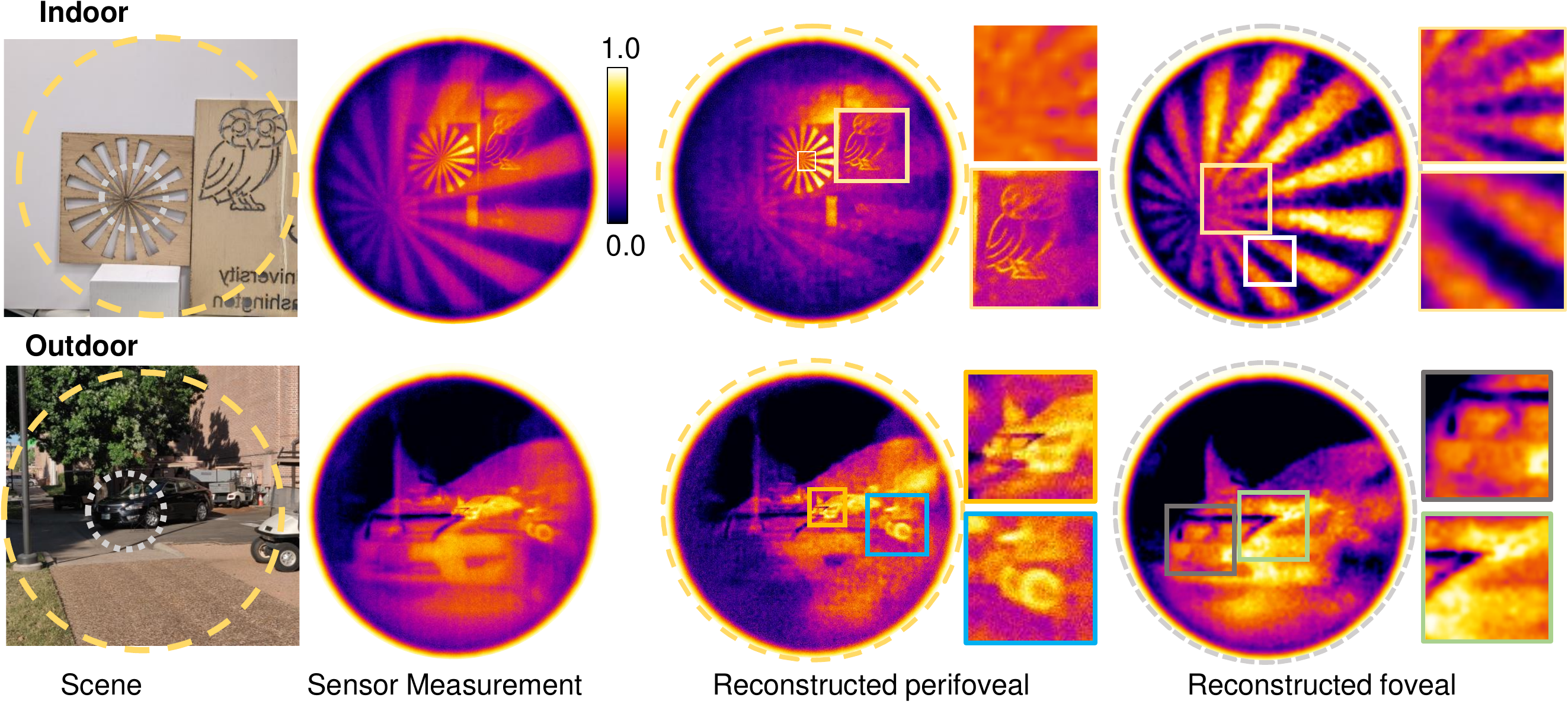}
	\caption{Snapshot recovery without polarizer. Our setup is capable of working without a rotating polarizer albeit with a small loss in reconstruction quality. The figure above shows an indoor and outdoor example. A single sensor measurement is shown with notable contributions from both lenses, followed by the reconstructed perifoveal and foveal images.}
	\label{fig:fig6}
\end{figure}

Figure~\ref{fig:fig6} shows an indoor example with a wooden stencil and an outdoor example with a parked car. The indoor scene has a perifoveal image consisting of the sector star target, as well as an owl, while the foveal image zooms into the sector star. The foveal image clearly shows the sector at high resolution. The sensor measurement for the outdoor scene has distinct contributions from both perifoveal (small car, and golf cart) and foveal (enlarged car) images. The reconstructed images show the surroundings in the perifoveal image including the wheels of a golf cart, while the foveal image shows the features of the car.
Note that there is some overlap in the reconstructions. The ill-posedness of the inverse problem produces artifacts in both perifoveal and foveal images. 
It is possible to remove such artifacts with more advanced computational approaches including a trained neural network~\cite{zhang2018single}, which we leave for future work.

%% file: disc.tex
We have demonstrated a first-of-its-kind optical system based on polarization-sensitive metalenses that enables foveation.
The optical setup along with a powerful computational backend is capable of imaging in the wild at real-time rates, opening up a wide range of possibilities in applications that require compact optical systems along with large field of view, and high spatial resolution. 

\begin{figure}
	\centering
	\includegraphics[width=\textwidth]{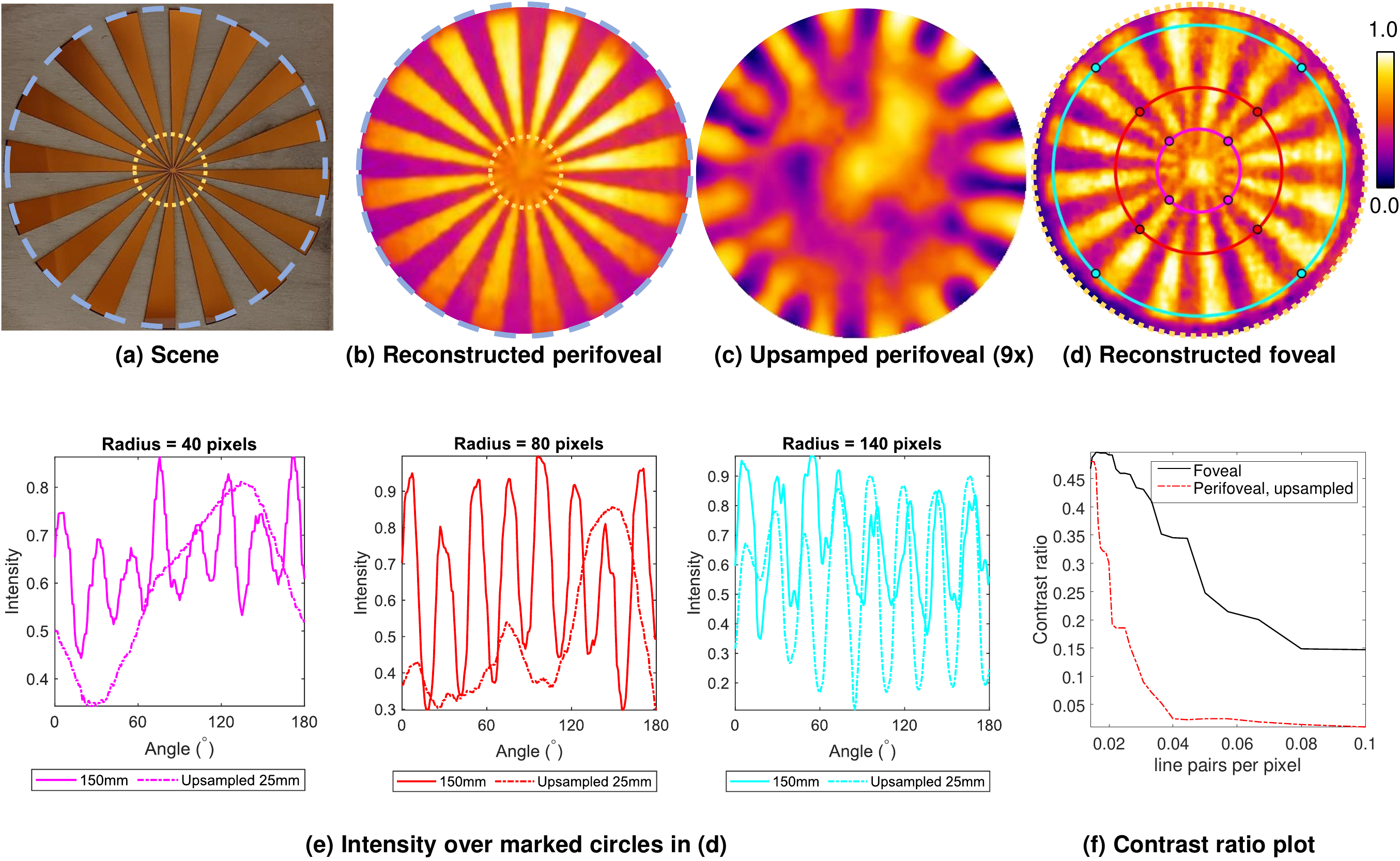}
	\caption{Resolution analysis. We captured a total of 100 images with the freely rotating polarizer, and reconstructed one perifoveal and one foveal image. The object was placed 300mm from the foveal lens, and hence the foveation expansion was $9:1$. (a) shows the sector star scene we captured with our optical system. (b) shows the reconstructed perifoveal image, (c) shows the upsampled perifoveal image, (d) shows the foveal image, (e) shows intensity over marked circles, and (f) shows contrast ratio as a function of line pairs per pixel. At the outer edge of the sector star, we notice that the perifoveal and foveal have similar contrast ratios. However on inner circles, the foveal image has significantly better contrast ratio. }
	\label{fig:fig2}
\end{figure}

\bpara{System analysis.} To quantify the performance of our optical system, we imaged a wooden cutout of the Seimen's sector target. The backside of the cutout consists of a thin chart paper heated by a portable heater to create temperature contrast. 
Figure~\ref{fig:fig2} shows the reconstructions with the 25mm and 150mm image as well as intensity profiles over circles with increasing radii. For large radii (140 pixels), the intensity profiles of the upsampled perifoveal image and the foveal image match. As the radius decreases (40, 80 pixels), the intensity profile for the foveal image continues to have high contrast, while the upsampled perifoveal image lacks any contrast. Figure~\ref{fig:fig2}(f) shows the contrast ratio as a function of line pairs per pixel (LPP). At very low lpp, the perifoveal and foveal images have similar contrast ratio. At larger radii, the foveal image has significantly higher contrast ratio, while the contrast of the perifoveal image reduces to a very small value.

\bpara{Limitations and future directions.}
Our metalens-based foveated imager currently produces sharp images only for a narrow range of wavelengths. This limitation is common across most metalenses~\cite{presutti2020focusing}.
However by co-optimizing the meta-optics and computational backend, the metalenses can be made to focus over a much broader range of wavelengths as shown in the visible~\cite{tseng2021neural}. Foveated imaging with broadband performance will constitute a promising future direction.
Our metalenses have moderate transmission efficiency (approximately 60\%), due to a lack of antireflection (AR) coating and absorption in the silicon substrate. This can be overcome with more advanced manufacturing procedures with AR coatings on the metalenses or using transparent materials in the LWIR range, such as As$_2$S$_3$ or GeSbSe thin-films on CaF$_2$ substrate~\cite{eggleton2011chalcogenide}.

%% file: methods.tex
\bpara{Metalens design.} To implement the polarization-dependent phase response in the two meta-optics we designed a scatterer with a polarization-dependent response using a rectangular footprint of each nanopost. We first calculated the phase and amplitude response for pillars using rigorous coupled wave analysis (RCWA)~\cite{liu2012s4}, assuming a pillar height of 10 $\mu$m, period of 4 $\mu$m and varying rectangular footprint defined by the sidewidths, x and y.
We then selected a scatterer that would have a phase response covering a range of 2$\pi$ for S-polarization, while not altering the phase for light with P-polarization. 
We note that meta-optics for imaging in LWIR have been demonstrated in recent times with polarization insensitivity~\cite{fan2018high}, broadband response~\cite{meem2019broadband}, large aperture~\cite{li2022largest}, and operation in ambient temperatures~\cite{huang2021long}. 
However, to the best of our knowledge, our system is the first to demonstrate results on polarization-sensitive metalenses at LWIR wavelengths.
Further details about the optimization procedure are available in section~\ref{sec:appendix1}.


\bpara{Metalens fabrication.} Each metalens was fabricated on a 500$\mu$m thick double-side polished silicon wafer, lightly doped with boron, giving a sheet resistivity of 10-20 $\Omega$-cm. We used direct-write lithography (Heidelberg DWL $66^+$) to define the location of the metalens aperture in a negative photoresist layer. A 240nm thick aluminum layer was deposited via electron beam evaporation (CHA solution) and lifted off to form the metal mask surrounding the designated aperture of the metalens, reducing the noise in the experiments. The metalens scatterer layout was aligned and patterned into the open circular aperture using direct-write lithography with a positive photoresist. We utilized deep reactive-ion etching (SPTS DRIE) to transfer the metalens pattern into the silicon layer with a scatterer depth of 10 $\mu$m and highly vertical sidewalls. 

\bpara{Optimization approach.}
%
Let $\textbf{x}^1 = I_1(x, y)$ and $\textbf{x}^2 = I_2(x, y)$ be the vector representation of perifoveal and foveal images respectively.
%
%
%

To regularize the inverse problem in eq.~\eqref{eq:optim}, we rely on the inherent regularization offered by convolutional neural networks and solve the following modified optimization problem,
\begin{align}
	\min_{\alpha_t, \theta_1, \theta_2} \sum_{t=t_1}^{t_2}\left\| \mathbf{y}_t -  \alpha_t \mathbf{x}^1 - (1-\alpha_t) \mathbf{x}^2\right\|^2 + \mathcal{R}( \mathbf{x}^1, \mathbf{x}^2)&\\
	\mathbf{x}^1 = \mathcal{N}_1(\mathbf{n}_1; \theta_1)&\\
	\mathbf{x}^2 = \mathcal{N}_2(\mathbf{n}_2; \theta_2)&,
\end{align}
where $\mathcal{N}_1$ and $\mathcal{N}_2$ are untrained neural networks, $\theta_1, \theta_2$ are weights of the neural networks to be optimized, and $\mathbf{n}_1$ and $\mathbf{n}_2$ are inputs with each entry drawn in a uniformly random manner.
This approach of optimizing for the weights of an untrained neural network instead of the image is similar in spirit to deep image prior~\cite{ulyanov2018deep}.
Utilizing two separate networks promotes dissimilarity between the two images $\mathbf{x}^1$ and $\mathbf{x}^2$, and hence results in high quality separation~\cite{gandelsman2019double}.
Further details about the network, as well as the regularization function $\mathcal{R}$, are available in section~\ref{sec:appendix1}.

%
For the snapshot case, given the perifoveal image $\mathbf{x}^1$ and the foveal image $\mathbf{x}^2$, we have
\begin{align}
	C\mathbf{x}^1 \approx D\mathbf{x}^2,
\end{align}
where $C$ is the cropping operator, and $D$ is the downsampling operator.
We then solve the modified optimization problem,
\begin{align}
	\min_{\mathbf{x}^1, \mathbf{x}^2} \| \mathbf{y} -  (\mathbf{x}^1 +  \mathbf{x}^2)\|^2 + \eta_1 \|C\mathbf{x}^1 - D\mathbf{x}^2\|^2 + \mathcal{R}( \mathbf{x}^1, \mathbf{x}^2),
\end{align}
where $\eta_1$ is the weight of penalty for similarity between the cropped perifoveal image and the downsampled foveal image.
Further details about the regularizer are available in section~\ref{sec:appendix2}.

\bpara{Experimental setup.} Our setup consists of two, four inch metalenses mounted on optomechanical systems (Thorlabs LMR4) on a rail system. A linear polarizer (Thorlabs WP25M-IRC) was mounted on a fast rotation stage (Thorlabs ELL14) and asynchronously rotated at 2Hz. 
Images in Fig.~\ref{fig:fig3}(a), (c), Fig.~\ref{fig:fig4}, Fig.~\ref{fig:fig5}, and Fig.~\ref{fig:fig2}  were captured with an Infratec VarioCAM HD 1800 camera, while the images in Fig.~\ref{fig:fig3}(b), (d), and Fig.~\ref{fig:fig6} were captured with a FLIR A655sc camera.
The Infratec VarioCAM HD 1800 camera was fitted with a $0.5\times$ relay lens, and the FLIR A655sc was fitted with a $0.3\times$ relay lens.
The relay system was required to overcome the mechanical constraints of the camera, which prevented us from placing the 25mm lens at the appropriate distance from the sensor. We found that the relay had no effect on the final quality of the acquired images, except for a magnification.
%
%
%
Further details are available in section~\ref{sec:appendix2}.

%% file: sup_algo.tex
\subsection{Network details}
\begin{figure}[!tt]
	\centering
	\includegraphics[width=\textwidth]{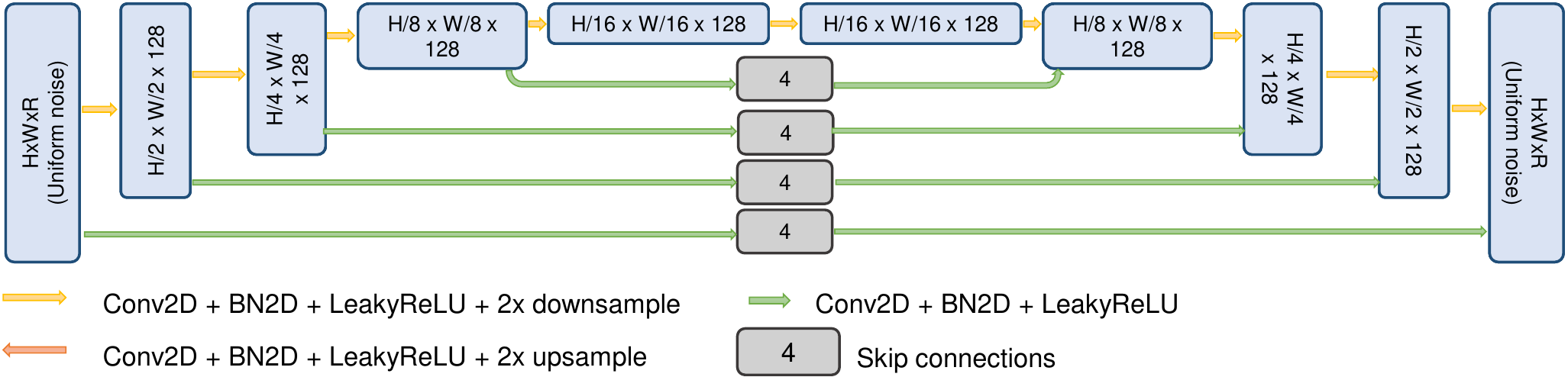}
	\caption{\textbf{Neural network architecture.} We used the skip net proposed in the deep image prior paper~\cite{ulyanov2018deep} with 128 features per each layer and a total of four layers. We used identical networks for both images with randomly initialized weights as well as inputs.}
	\label{fig:net}
\end{figure}

The optimization function used for recovering the foveal and perifoveal images is given by
\begin{align}
	\min_{\alpha_t, \theta_1, \theta_2} \sum_{t=t_1}^{t_2}\| \mathbf{y}_t -  \alpha_t \mathbf{x}^1 - (1-\alpha_t) \mathbf{x}^2\|^2 + \mathcal{R}( \mathbf{x}^1, \mathbf{x}^2)&\\
	\mathbf{x}^1 = \mathcal{N}_1(\mathbf{n}_1; \theta_1)&\\
	\mathbf{x}^2 = \mathcal{N}_2(\mathbf{n}_2; \theta_2)&.
\end{align}
Inspired by the deep image prior (DIP) work, we chose both the networks to be UNet~\cite{ronneberger2015u} with skip connections. 
The specific network structure produces outputs that resemble images, and hence is robust to noise. Figure~\ref{fig:net} shows the network architecture for each image.
Prior work by Gandelsman et.\ al.~\cite{gandelsman2019double} has shown that utilizing two separate networks to produce a  ``double DIP'' representation encourages dissimilar images.

\subsection{Fixed pattern noise}
Microbolometer cameras suffer from slowly varying fixed pattern noise and non-uniformities due to manufacturing defects and internal thermodynamics.
Recent works~\cite{he2018single} identified that such non-uniformities tend to be low rank. The works by \cite{saragadam2021thermal} have shown that these non-uniformities do not change over a short duration of time and hence can be modeled as constants.
Inspired by this work, we introduce non-uniformity as another variable to be optimized, giving us,
\begin{align}
	\min_{\alpha_t, \theta_1, \theta_2} \sum_{t=t_1}^{t_2}\| \mathbf{y}_t -  \alpha_t \mathbf{x}^1 - (1-\alpha_t) \mathbf{x}^2+\bfn_\text{NU}\|^2 + \mathcal{R}( \mathbf{x}^1, \mathbf{x}^2)&\\
	\mathbf{x}^1 = \mathcal{N}_1(\mathbf{n}_1; \theta_1)&\\
	\mathbf{x}^2 = \mathcal{N}_2(\mathbf{n}_2; \theta_2)&\\
	\text{Rank}(\bfn_\text{NU}) = K&,
\end{align}
where $\bfn_\text{NU}$ is the non-uniform noise to be corrected. We chose the rank to be 20 for all our experiments. Further, we constrained the entries of $\bfn_\text{NU}$ to lie between $-0.1$ and $0.1$. 
Some of the recovered noise images are visualized in the upcoming experiments section.

\subsection{Regularization}
The regularization function $\mathcal{R}(\bfx_1, \bfx_2)$ is,
\begin{align}
	\mathcal{R}(\bfx_1, \bfx_2) &= \eta_\text{TV}(\|\nabla\bfx_1\|_1 + \|\nabla\bfx_2\|_1) + \eta_\text{excl.}\mathcal{L}_\text{excl.}(\bfx_1, \bfx_2),
\end{align}
where $\|\cdot\|_1$ is the total variation (TV) loss, and $\mathcal{L}_\text{excl.}$ is the gradient exclusion loss~\cite{zhang2018single} that encourages $\bfx_1$ and $\bfx_2$ to be different.
The penalty terms $\eta_\text{TV}$ was set to $0.01$, and $\eta_\text{excl.}$ was set to $0.01$. 

\begin{figure}[!tt]
	\centering
	\includegraphics[width=\textwidth]{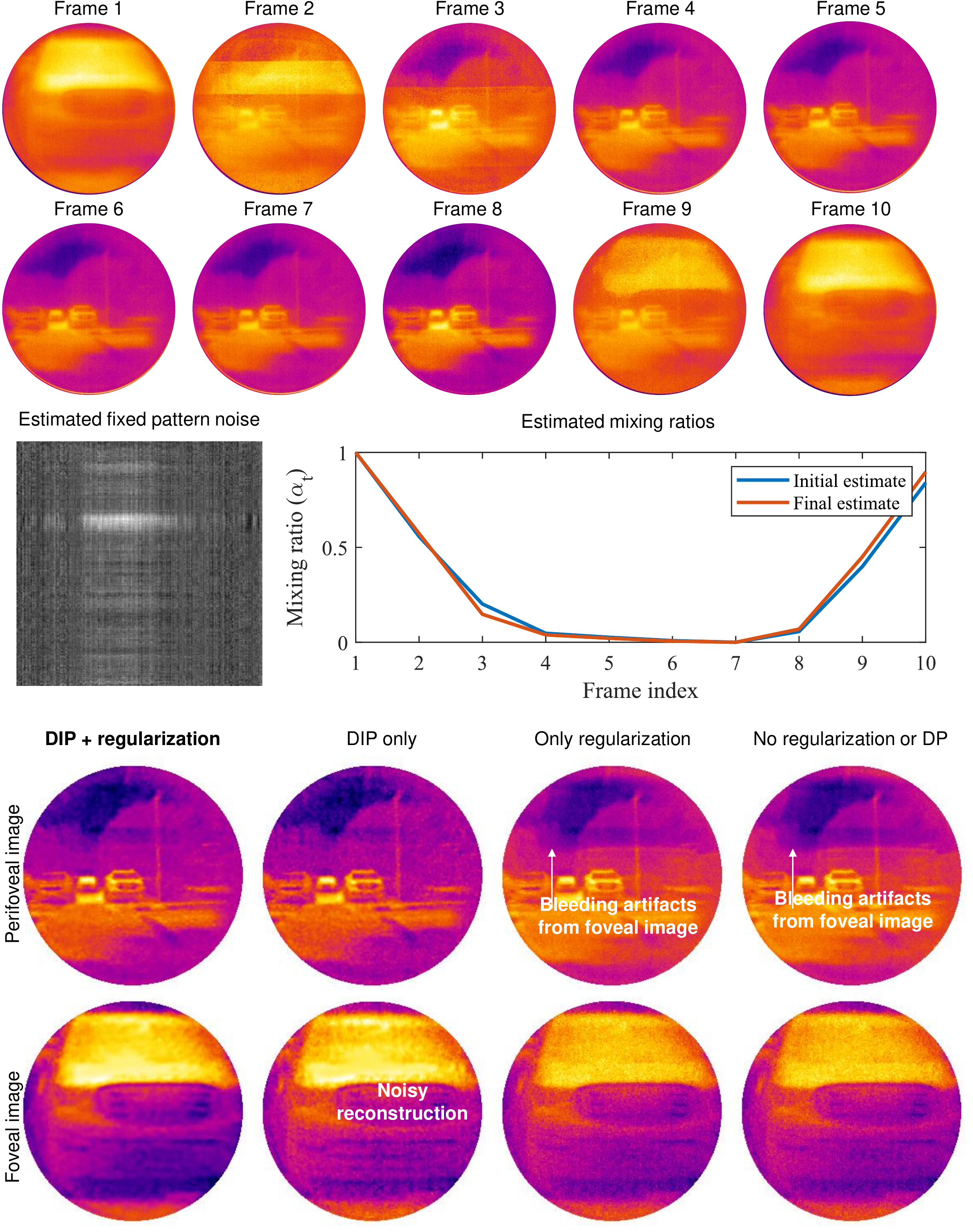}
	\caption{Multiframe recovery example. The figure above shows input images, the recovered 25mm and 150mm focal length images, as well as mixing ratios for a simulated example. The bottom row also shows reconstructions for various regularization options that showcase the benefits of utilizing a double DIP, as well as the inclusion of exclusion loss.}
	\label{fig:sim_multi}
\end{figure}

\subsection{Effect of initial mixing ratios}
Since the reconstruction problem is highly non-convex, the initialization of each variable plays an important role in recover.
We found that initializing the weights with mean intensity of each video frame lead to a faster convergence and accurate estimate of true mixing ratios.
Figure~\ref{fig:sim_multi} shows an example on simulated images with and without regularization. The addition of a double DIP, as well as the gradient exclusion loss considerably improve the quality of reconstruction.

\begin{figure}[!tt]
	\centering
	\includegraphics[width=\textwidth]{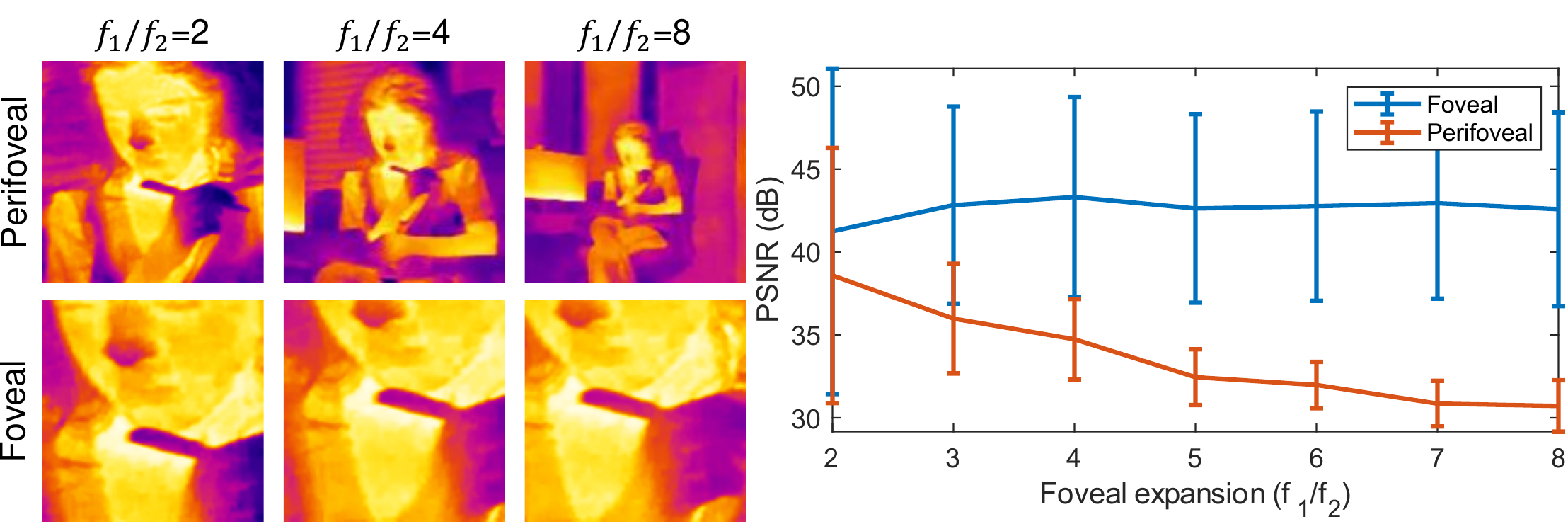}
	\caption{Accuracy as a function of foveal expansion. We demonstrated lenses with a foveal expansions of $f_1/f_2=6$ but other foveal expansions are easy to implement. The figure above shows reconstructions with foveal expansions of 2, 4, and 8, as well as accuracy as a function of focal ratio, averaged over three images for a fixed foveal focal length. We notice that the accuracy of the foveal image is nearly constant, while that of perifoveal image reduces with increasing foveal expansion. The reason for this decrease is the increasing texture in the perifoveal image.}
	\label{fig:focal_ratio}
\end{figure}

\subsection{Effect of focal ratio}
We showed results with a foveal focal length of 150mm and perifoveal focal length of 25mm, leading to a foveal expansion of $6:1$, which was largely inspired by relative acuity of human eye.
It is however possible to build systems with arbitrary focal ratios.
To evaluate the accuracy of estimate with various foveal expansions, we simulated reconstruction with foveal expansions from 2 to 8 in steps of 1.
For each case, we simulated a capture of 10 images with varying mixing ratios.
Figure~\ref{fig:focal_ratio} shows images with each foveal expansion, as well as accuracy (PSNR) across configurations averaged over 3 simulated examples for a fixed foveal focal length.
The reconstruction accuracy stays nearly the same for foveal image, since the focal length was kept fixed.
In contrast, the PSNR for perifoveal image reduces with increasing focal ratio, primarily due to increasing texture.
More importantly, our approach can be gracefully extended to nearly any foveal expansion as required by the end application.

\subsection{Comparison against other foveated systems}
\begin{figure}[!tt]
	\centering
	\includegraphics[width=\textwidth]{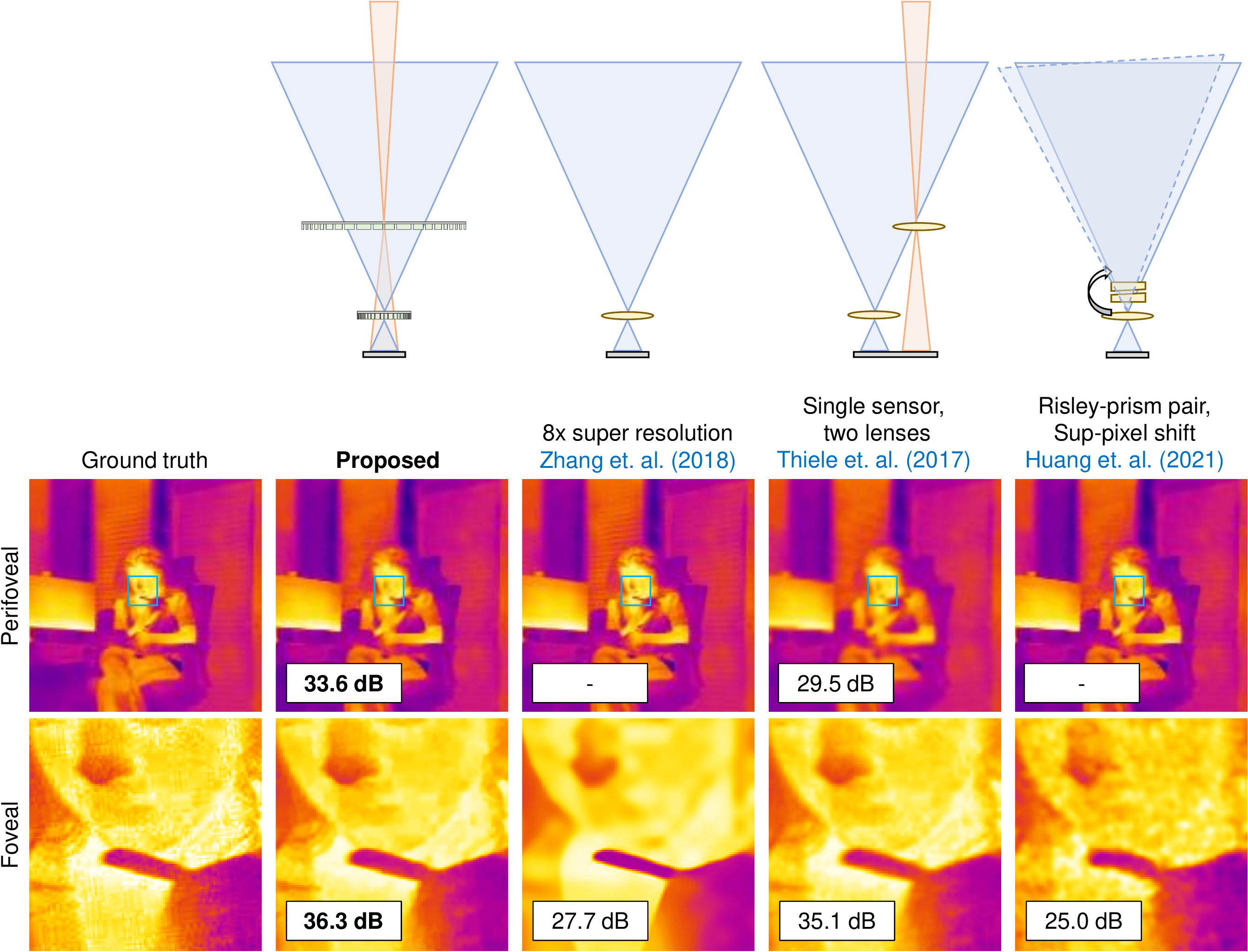}
	\caption{Comparison against other foveated systems. There are very few optical systems for dedicated foveated imaging. A software-based super resolution approach~\cite{zhang2018image} requires a single camera, but leads to smooth results. Solutions including a single sensor but two separate lenses such as that by Thiele et al.\ \cite{thiele20173d} are closest to our approach, but lead to a spatial resolution that is half that of the sensor resolution. Techniques that use a moving optic such as a Risley-prism~\cite{huang2021flexible} leverage sub-pixel shift to super resolve the image, but requires several images to construct a single frame, precluding video-rate recovery. Across the board, our approach achieves best results, both qualitatively and quantitatively.}
	\label{fig:compare}
\end{figure}

Though there are some works that utilize polarization for foveated displays~\cite{yin2022foveated,yoo2020foveated}, there are very few imaging systems that enable double focal length foveation with a single sensor.
Ude et al.\ \cite{ude2006foveated} showed that foveation is useful for robotic manipulation but used two sensors with two different focal lengths.
Carles et al.\ \cite{carles2016multi} leveraged sub-pixel shifts in the central area of the sensor to obtain super resolution, but used a large array of camera sensors.
To the best of our knowledge, two works are close to the proposed approach in the paper.
Thiele et al.\ \citep{thiele20173d} utilized two smaller lenses on top of a CMOS sensor to obtain foveation.
Huang et al.\ \citep{huang2021flexible} utilized a rotating risley prism pair to obtain several measurements with sub-pixel shifts.
We compare our approach against these two approaches, as well as a pure algorithm-only super resolution using channel attention-based neural networks~\cite{zhang2018image}.
Figure~\ref{fig:compare} compares various approaches visually and quantitatively in terms of PSNR and SSIM for an example image.
Across the board, we observe that our polarization multiplexed approach achieves superior performance both visually and quantitatively.

\subsection{Single image recovery}
For snapshot case, we have a single measurement of the form,
\begin{align}
    \bfy =  \alpha\bfx^1 + (1-\alpha)\bfx^2,
\end{align}
By design, the downsampled version of $\bfx^1$ is (approximately) equal to the center crop of $\bfx^2$ (assuming $f_1 > f_2$). This gives us,
\begin{align}
    D_{f_1/f_2} \bfx^1 = C \bfx^2,
\end{align}
where $D_n$ is a downsampling operator by $n$ times, and $C$ is the cropping operator.

The specific optimization objective for recovering the two images,
\begin{align}
    \min_{\theta_1, \theta_2} \| \bfy - (\alpha \bfx^1 + (1-\alpha) \bfx^2)\|^2 &+ \eta_1\|D_{f_1 / f_2}\bfx^1 - C\bfx^2\|^2 + \eta_2 \mathcal{L}_\text{excl}(\bfx^1, \bfx^2)\\
    &\bfx^1 = \mathcal{N}_1 (\bfn_1; \theta_1)\\
    &\bfx^2 = \mathcal{N}_2 (\bfn_2; \theta_2).
\end{align}

Prior work including \cite{ulyanov2018deep} and \cite{gandelsman2019double} modeled $\bfn_1, \bfn_2$ as uniform random variables. However, we found that the following initialization was critical to a successful recovery of the two images,
\begin{align}
    \bfn_1 &= D^\top_{f_1/f_2}\bfy \\
    \bfn_2 &= \bfy - \bfn_1.
\end{align}
The key idea is that an upsampled version of the observation $\bfy_1$ resembles the high resolution image $\bfx_1$, and hence is a good initial estimate.

\begin{figure}[!tt]
    \centering
    \includegraphics[width=\textwidth]{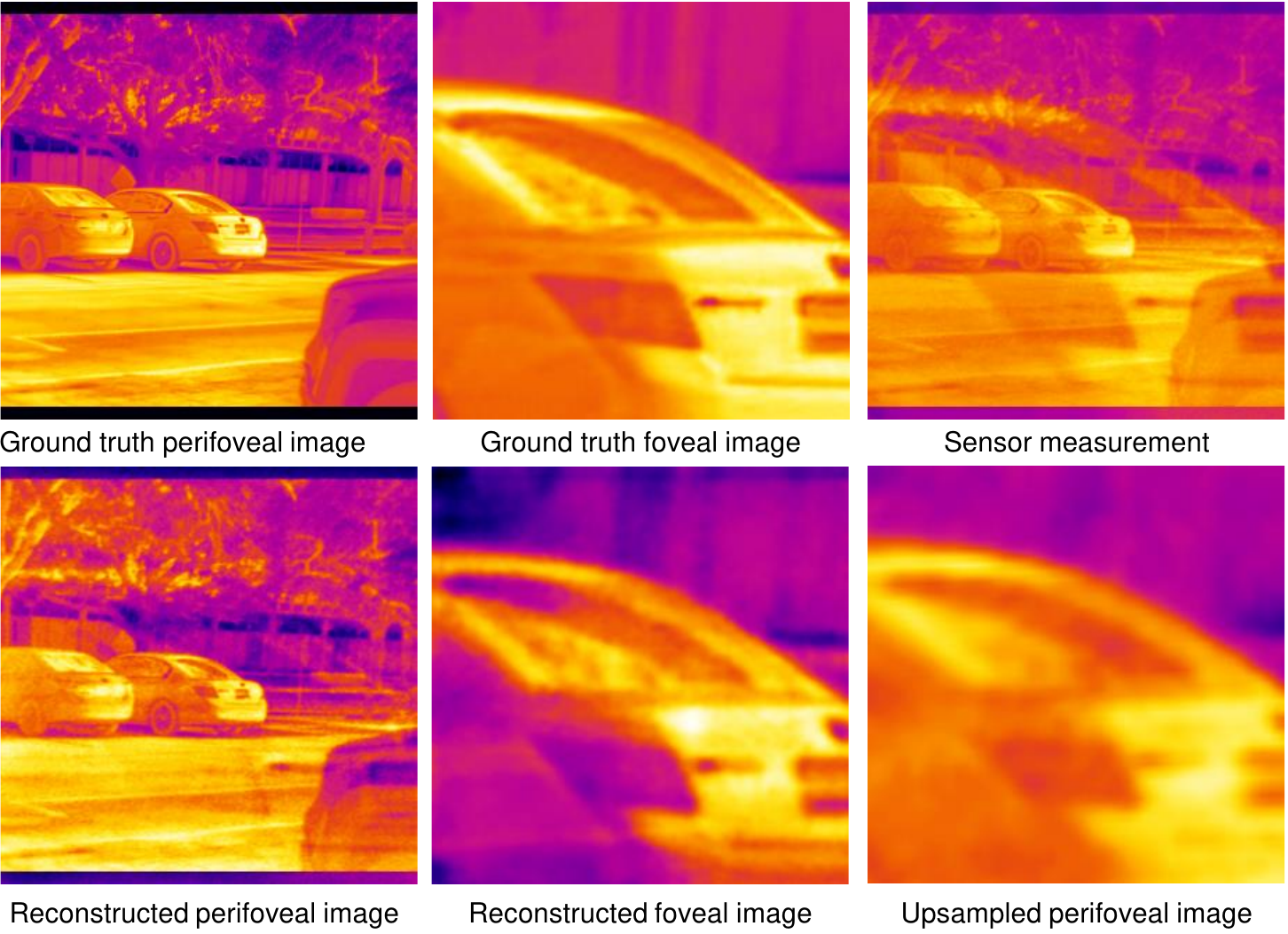}
    \caption{Simulated reconstruction. (a) shows image simulated with large FoV lens, and (b) shows the image simulated with the high resolution lens. (c) is the simulated sum image. (d) and (f) are the reconstructed separate images, while (g) shows an upsampled version of (d). Our image separation algorithm results in high quality reconstruction.}
    \label{fig:sim1}
\end{figure}

\begin{figure}[!tt]
    \centering
    \includegraphics[width=\textwidth]{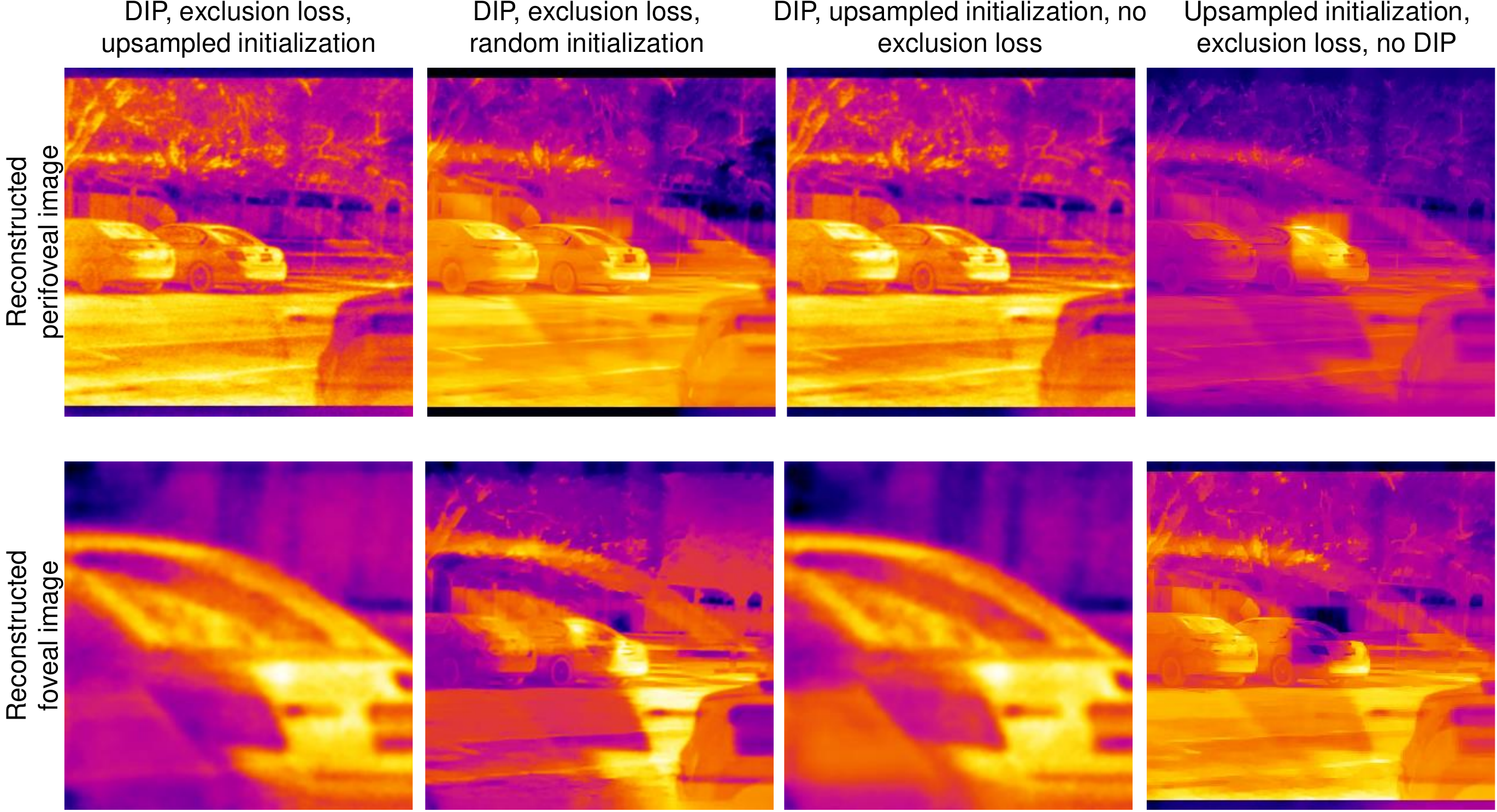}
    \caption{Effect of each component in snapshot recovery algorithm. The figure above compares the effect of initialization, exclusion loss, and usage of deep prior. We observe that initialization and deep prior have the strongest effect on separation.}
    \label{fig:ablation}
\end{figure}

Figure \ref{fig:sim1} demonstrates reconstruction with our approach on a simulated image, and Fig.~\ref{fig:ablation} evaluates the effect of each component in the recovery algorithm on the final separation.
As with multi-frame recovery, we observe that the initialization, deep prior, and the exclusion loss work in tandem to produce a high quality reconstruction.

%% file: sup_experiments.tex
\begin{figure}[!tt]
	\centering
	\includegraphics[width=0.7\textwidth]{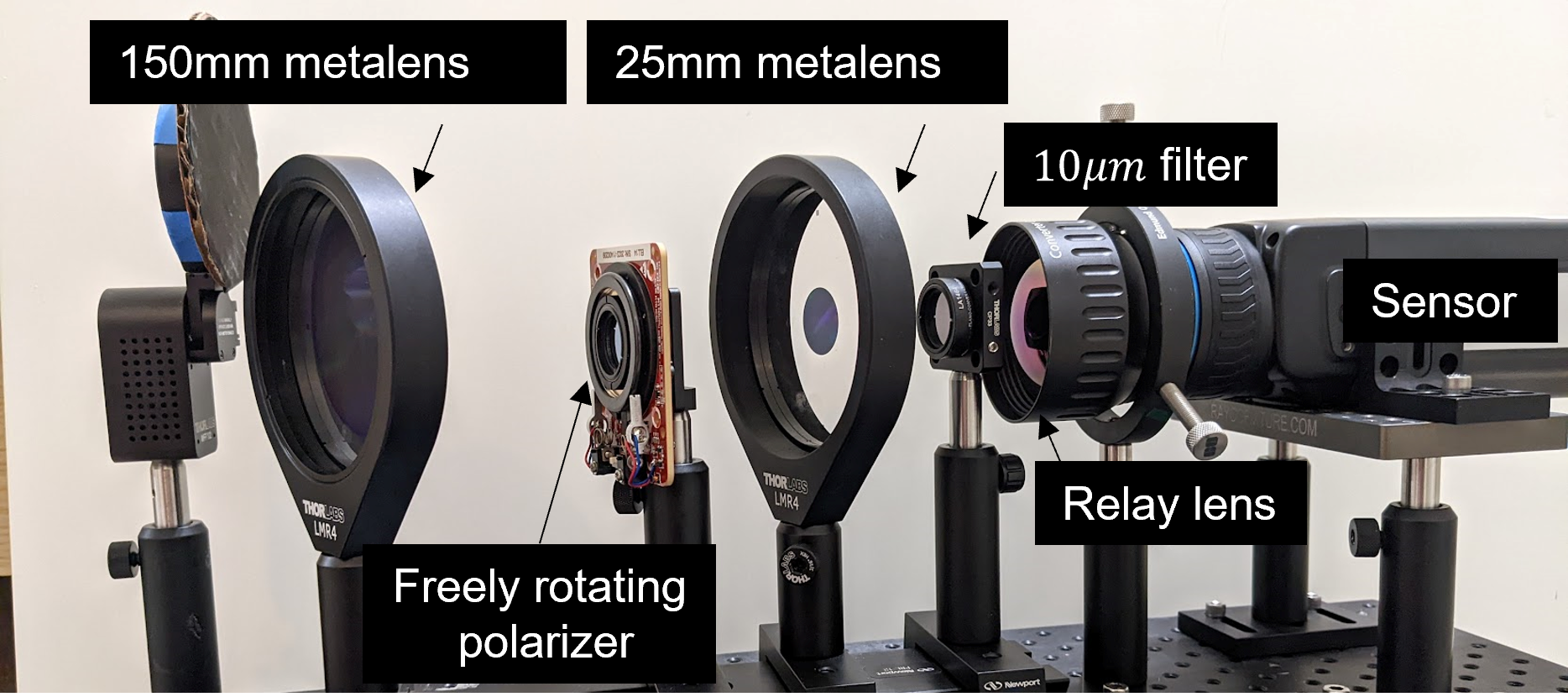}
	\caption{Lab prototype. Our setup consists of two metalenses of 150mm (F/2) and 25mm (F/1) focal lengths, a freely rotating polarizer, and a $10\mu$m narrowband filter (500nm bandwidth). A relay lens was used between the lenses and the sensor due to mechanical difficulties of placing the 25mm metalens sufficiently close to the sensor. }
	\label{fig:setup}
\end{figure}

\subsection{Lab prototype}
Figure~\ref{fig:setup} shows our lab prototype consisting of two metalenses, a freely rotating polarizer, a narrowband filter, and relay lens.
We used a FLIR A655sc longwave infrared (LWIR) camera to capture raw images.
Additionally, we placed black shutter (Acktar Black metal velvet) in front of the 150mm metalens to account for stray light and internal reflections (narcissus effect).
We captured images from the camera using the Spinnaker Python software development kit (SDK) from FLIR.
The polarizer and the flip mount were controlled with python software as well.

\begin{figure}[!tt]
\centering
\includegraphics[width=\textwidth]{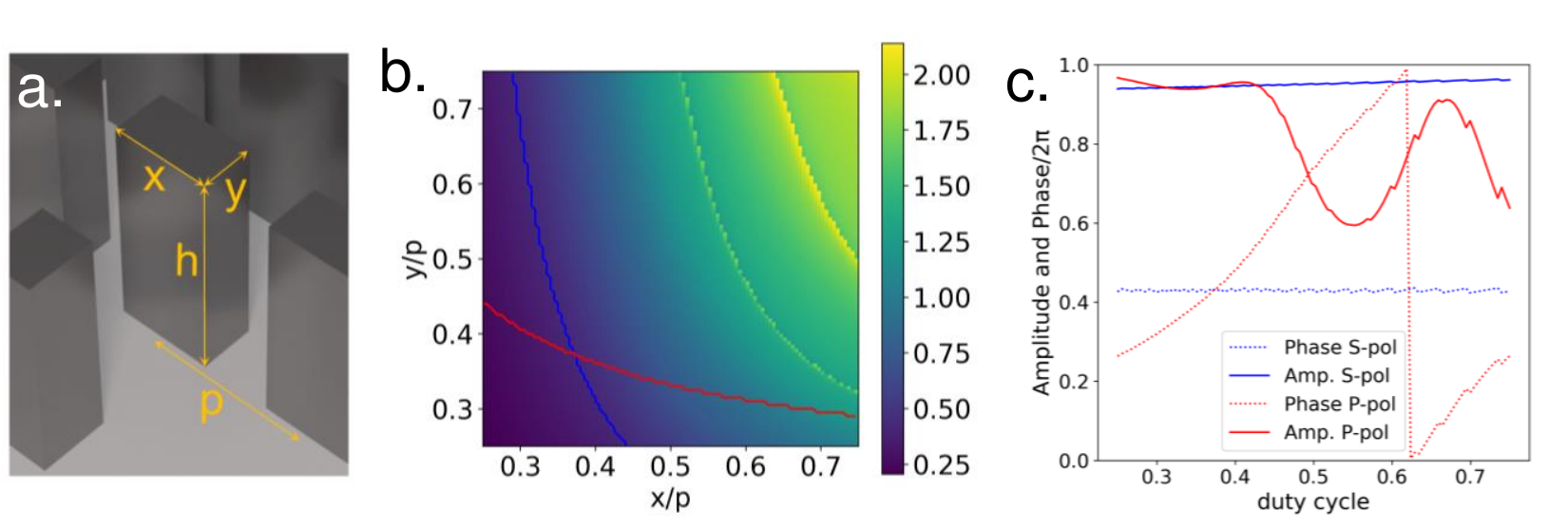}
\caption{Design and selection of the scatterer. a) Schematic of the scatterer dimensions, where p is the period, h is the height, x and y are the scatterer side widths.  b) Map of the phase response in dependence of the duty cycles x/p and y/p. The red and blue line represent the response of the selected scatterer to orthogonal polarizations. c) Amplitude and phase response to orthogonal polarizations.    }
	\label{fig:fig7}
\end{figure}

\subsection{Metalens design}
To implement the polarization dependent phase response in the two meta-optics, we designed scatterer with a polarization dependent response using a rectangular footprint of each nanopost. We first calculated the phase and amplitude response for pillars using rigorous coupled wave analysis (RCWA)~\cite{liu2012s4}, assuming a pillar height of 10 $\mu$m, period of 4 $\mu$m and varying rectangular footprint defined by the side widths, x and y.
A map of the phase response for light with S-polarization in dependence of the respective duty cycles (x/p and y/p) is shown in (Figure~\ref{fig:fig7}b).
We then selected scatterer that would have a phase response covering a range of 2$\pi$ for S-polarization (red curve), while not altering the phase for light with P-polarization (blue curve). 
The extracted phase response is plotted in (Figure~\ref{fig:fig7}c). 

\subsection{Real Experiments}
\begin{figure}[!tt]
	\includegraphics[width=\textwidth]{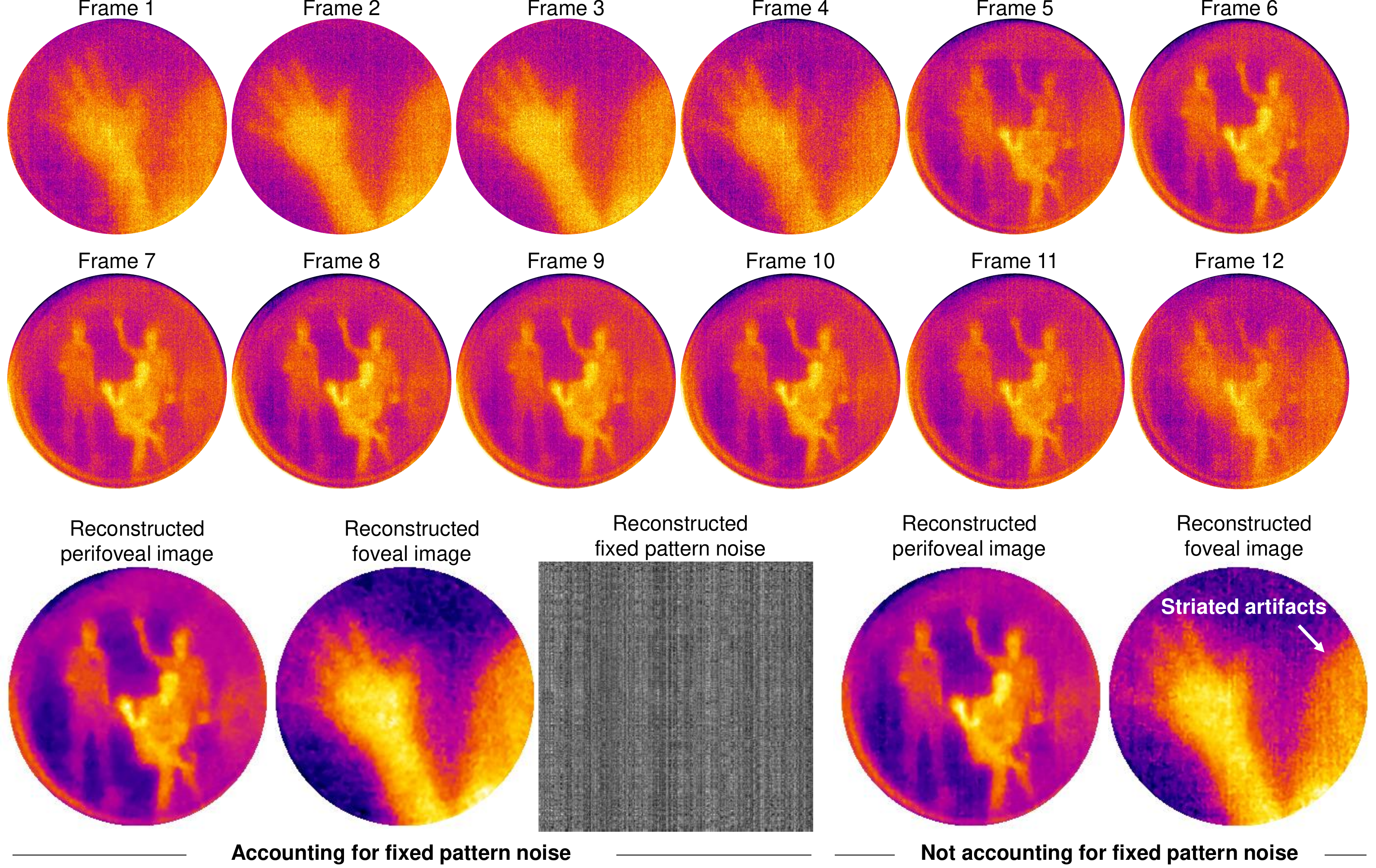}
	\caption{Effect of fixed pattern noise. Images captured with low signal levels tend to have strong fixed pattern noise. Accounting for them produces sharper images. In contrast, not accounting for the fixed pattern noise produces striated artifacts that correlate with the noise.}
	\label{fig:fpn}
\end{figure}

%

%

\subsubsection{Effect of including fixed pattern noise}
To demonstrate the advantages of accounting for the fixed pattern noise, we compared the reconstructions with and without the fixed pattern noise.
Figure~\ref{fig:fpn} shows a side-by-side comparison of the reconstructions for a noisy measurement.
Not including the fixed pattern noise produces strong column artifacts that correlate with the noise pattern.
In contrast, recovery that accounts for the fixed pattern noisy is considerably sharper and devoid of any such artifacts.

\subsubsection{Optimal number of images for video reconstruction}
\begin{figure}[!tt]
	\includegraphics[width=\textwidth]{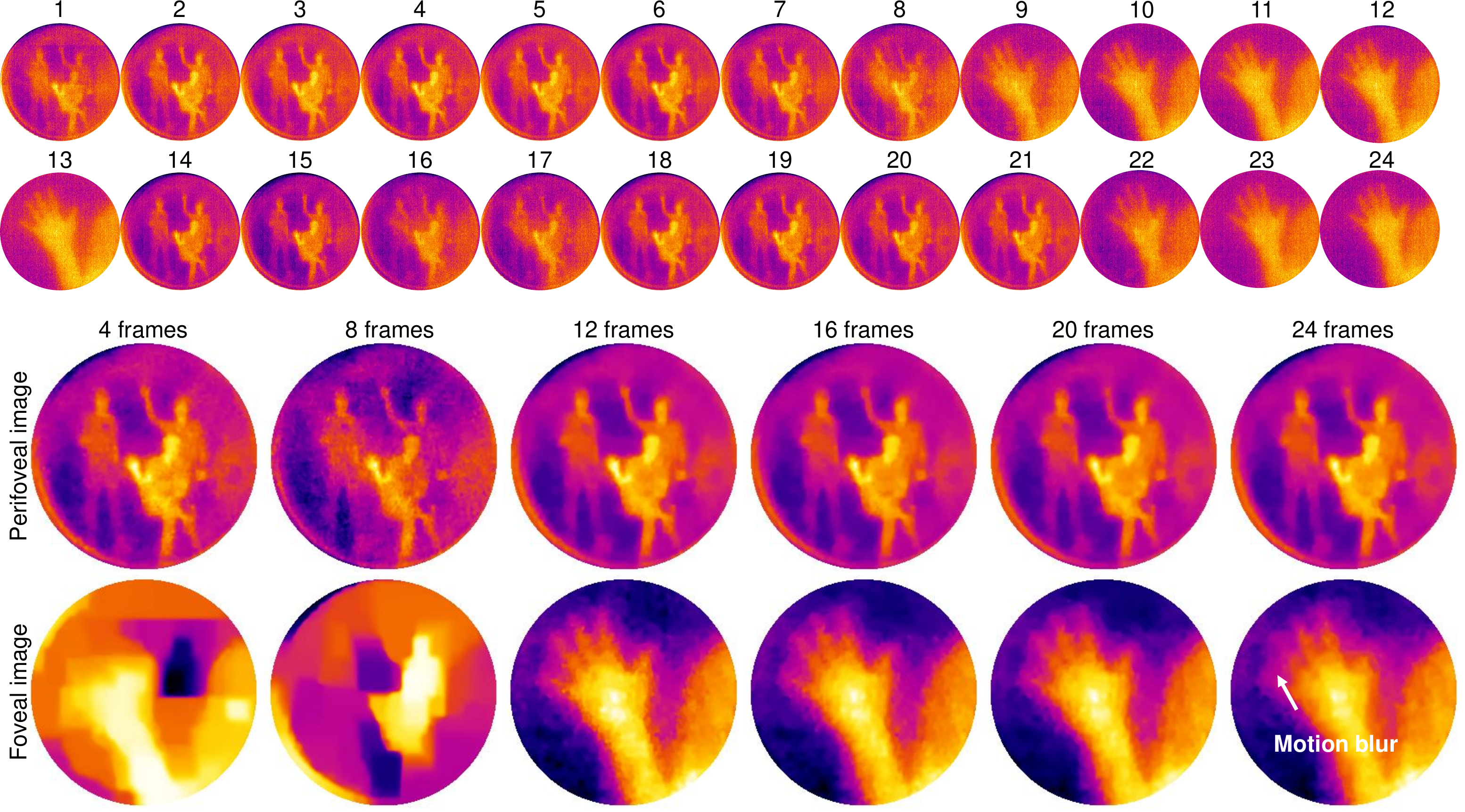}
	\caption{Effect of number of frames. Reconstruction with very few results in poor separation as the input images do not have sufficient information from foveal and perifoveal lenses. In contrast, using a lot of frames results in motion blur. In our experiments, we found 12 - 16 frames to be optimal for getting high quality reconstruction.}
	\label{fig:nframes}
\end{figure}

The optimal number of images for reconstructing each frame of a video depends on the speed of polarizer, and the motion in the scene.
To empirically verify the effect of number of images, we varied the number of inputs for the hand waving scene from 4 to 24 in steps of 4.
Figure~\ref{fig:nframes} shows a comparison for varying number of images.
For fewer than 12 images, the reconstruction had significant overlap between the 150mm image and the 25mm image. 
In contrast, if the images were more than 20, the reconstructions had strong motion blurring.
We found the optimal number of images to be between 8 and 12, as it corresponded to a full period of the polarizer rotating at 2Hz, while frames were captured at 12Hz.
We do note that a rapidly changing scene would cause motion artifacts. 
This can be readily remedied with faster polarizers, or with sensors that are equipped with an array of linear polarizers.

\subsubsection{Real-time reconstruction}
The video results in the main paper were reconstructed with a deep image prior to obtain high quality results.
However, the optimization procedure takes up to thousand gradient descent steps, precluding real-time reconstruction.
It is however possible to obtain real-time reconstruction at the cost of noisy reconstructions.
To do so, we need to know the mixing ratios apriori. This is very much possible, as the rotating polarizer can be fully synchronized with the camera exposure duration.
In such conditions, estimating the foveal and perifoveal image involves solving a simple and computationally inexpensive least squares problem,
\begin{align}
	\min_{\bfx^1, \bfx^2} \sum_{t=t_1}^{t_2}\| \mathbf{y}_t -  \alpha_t \mathbf{x}^1 - (1-\alpha_t) \mathbf{x}^2\|^2.
\end{align}
The solution to the above least squares has a simple closed form given by,
\begin{align}
	\widehat{\bfx^2} &= \frac{a_{12}\bfs_1 - a_{11}\bfs_2}{a^2_{12}-a_{11}a_{22}}\\
	\widehat{\bfx^1} &= \frac{\bfs_1 -a_{12}\widehat{\bfx^2}}{a_{11}}\\
	\bfs_1 = \sum_{t=t_1}^{t_2}\alpha_t\bfy_t, &\quad
	\bfs_2 = \sum_{t=t_1}^{t_2}(1-\alpha_t\bfy_t)\\
	a_{11} =\sum_{t=t_1}^{t_2}\alpha_t^2,\quad
	a_{12} &=\sum_{t=t_1}^{t_2}\alpha_t(1-\alpha_t),\quad
	a_{22} =\sum_{t=t_1}^{t_2}(1-\alpha_t)^2
\end{align}
\begin{figure}[!tt]
	\includegraphics[width=\textwidth]{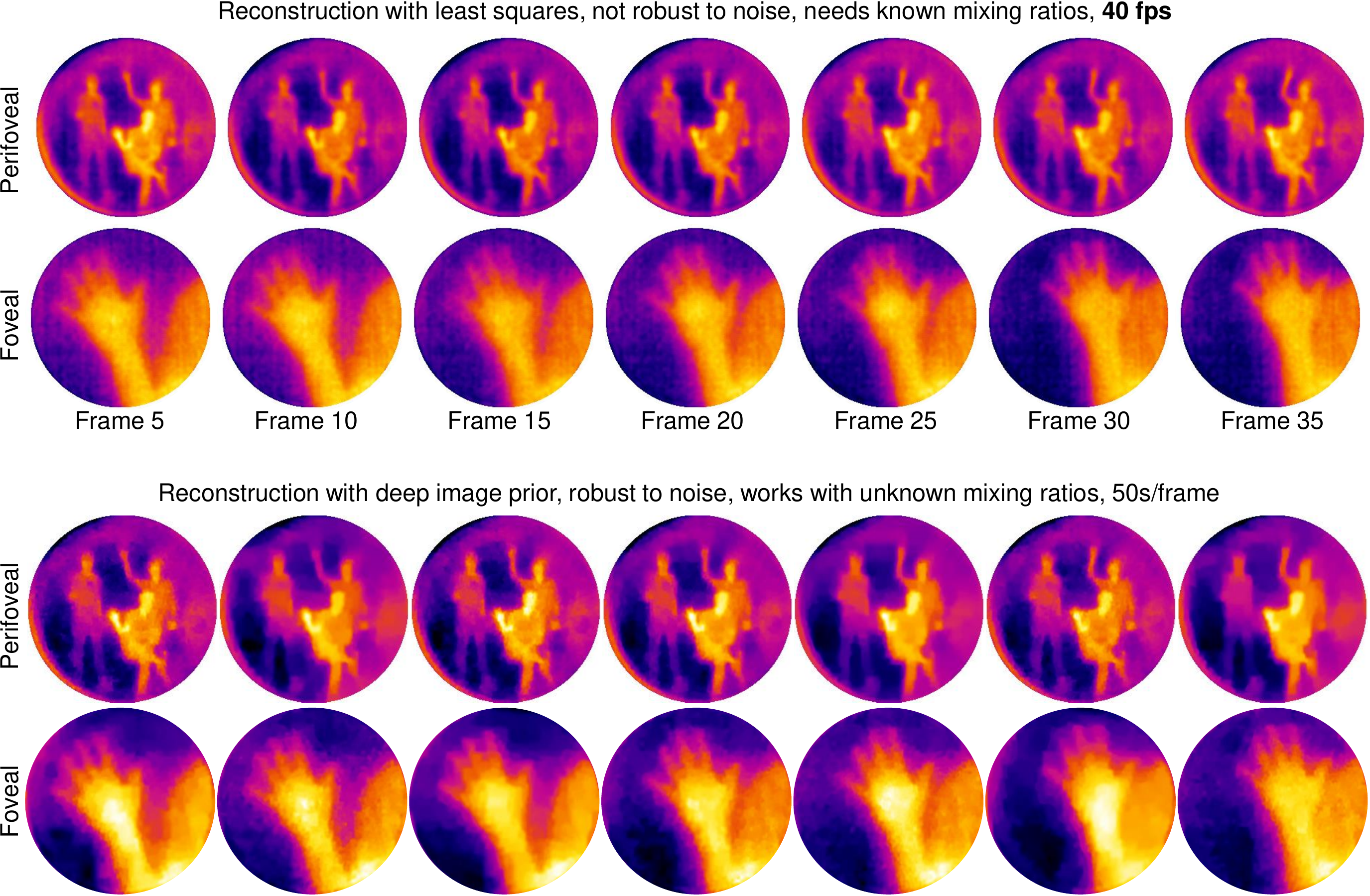}
	\caption{Real-time reconstruction. If the mixing ratios are known apriori, the foveal and perifoveal images can be reconstructed by solving a simple least squares expression. Such an approach is real-time (top row) but at the cost of increased noise in reconstruction. In contrast, a deep prior-based reconstruction (bottom) is robust to noise, can work with unknown mixing ratios, but requires up to 50 seconds for each video frame.}
	\label{fig:lstsq}
\end{figure}
To test this, we reconstructed the hand waving scene in the main paper with known mixing ratios.
Figure~\ref{fig:lstsq} shows a comparison of reconstruction with deep prior (robust but not real-time) with unknown mixing ratios, and least squares (real-time but not robust) with known mixing ratios. We performed an additional bilateral filtering to reduce the noise in least squares-based reconstruction.
We notice that the least squares solution is noisier compared to deep prior results, but we can obtain results at 40 fps for $275\times280$ sized image.
Future work will focus on training neural networks that can robustly separate the two images in real time with efficient GPU-based implementations.
\begin{figure}[!tt]
	\includegraphics[width=\textwidth]{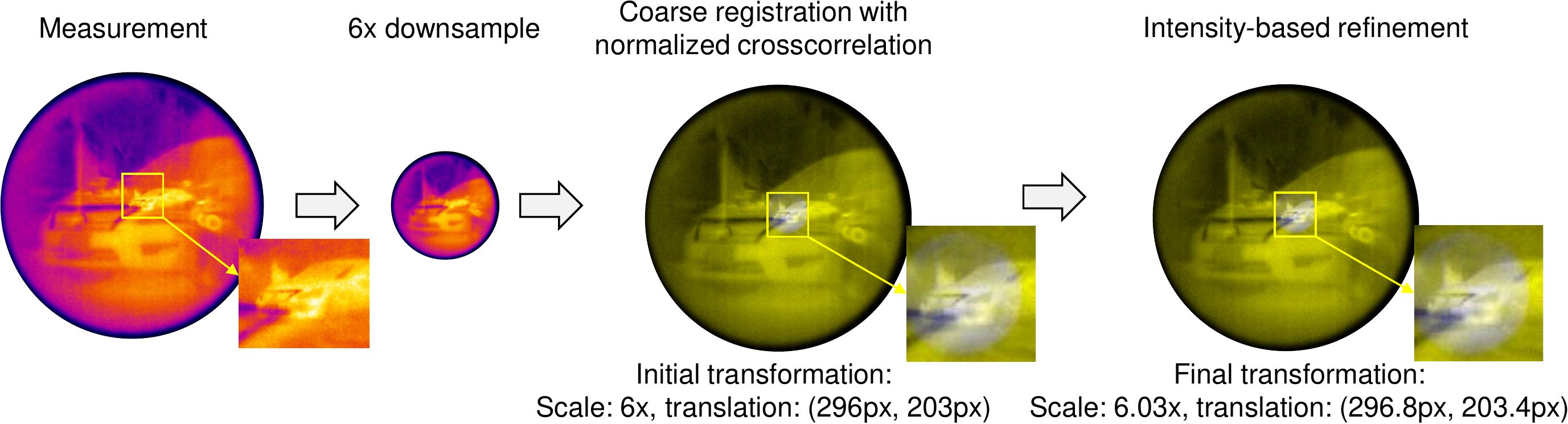}
	\caption{Registration for snapshot recovery. We follow a two step registration process. In the first step, we assume the scale to be $1:6$ and estimate only the translation using normalized crosscorrelation. In the second step, we refine both scale and translation using intensity-based registration. The two images on the right show overlap of the measurement (red and green channels) and registered downsampled images (blue channel).}
	\label{fig:reg}
\end{figure}
\subsubsection{Calibration for snapshot recovery}
Recovery from a single image (without polarizer) requires knowledge of the downsampling operator $D$ and cropping operator $C$. 
We assume that the 150mm image and the 25mm image are related by a simple rigid transformation consisting of scaling and translation.
It is possible to perform a one-time calibration of the system using standard geometric targets such as a checker board, however, since the lenses were not fixed in place, each capture had a unique set of transformation parameters.
To estimate the parameters for each scene, we relied on a two stage approach. 
First, we assumed the scale to be fixed and equal to the ratio of the focal lengths.
We then estimated the translation using normalized cross correlation between the captured combination image, and its downsampled version.
We then refined the scale and translation with an intensity-based registration approach.
Figure~\ref{fig:reg} visualizes the calibration procedure using our two stage approach.